\documentclass[twocolumn,trackchanges,twocolappendix]{aastex701}
\received{16 July, 2025}

\submitjournal{\apjs}

\usepackage{amsmath,environ}
\usepackage{listings}
\usepackage{float}
\usepackage{color}
\usepackage{mathtools}
\usepackage{rotating}

\usepackage{lineno}
\linenumbers

\newcommand{\halfwdth}{0.475\textwidth}

\newcommand{\hii}{H~{\sc ii}}
\newcommand{\hi}{H~{\sc i}}

\newcommand{\hneut}{H\textsuperscript{0}}

\newcommand{\ci}[1]{[C~{\sc i}]{#1}} 
\newcommand{\cii}[1]{[C~{\sc ii}]{#1}} 
\newcommand{\nii}[1]{[N~{\sc ii}]{#1}} 
\newcommand{\niii}[1]{[N~{\sc iii}]{#1}} 
\newcommand{\oi}[1]{[O~{\sc i}]{#1}} 
\newcommand{\oiii}[1]{[O~{\sc iii}]{#1}} 
\newcommand{\neii}[1]{[Ne~{\sc ii}]{#1}} 
\newcommand{\neiii}[1]{[Ne~{\sc iii}]{#1}} 

\newcommand{\ha}{H$\alpha$}

\newcommand{\niio}[1]{[N~{\sc ii}]$\lambda${#1}} 
\newcommand{\oio}[1]{[O~{\sc i}]$\lambda${#1}} 
\newcommand{\oiiio}[1]{[O~{\sc iii}]$\lambda${#1}} 
\newcommand{\siio}[1]{[S~{\sc ii}]$\lambda${#1}} 

\newcommand{\hp}{H\textsuperscript{+}}
\newcommand{\cp}{C\textsuperscript{+}}
\newcommand{\op}{O\textsuperscript{+}}
\newcommand{\opp}{O\textsuperscript{2+}}
\newcommand{\np}{N\textsuperscript{+}}
\newcommand{\npp}{N\textsuperscript{2+}}
\newcommand{\spion}{S\textsuperscript{+}}

\newcommand{\oh}{$\log$\,(O/H)}
\newcommand{\no}{$\log$\,(N/O)}

\newcommand{\edens}{$n_e$}
\newcommand{\hdens}{$n_\mathrm{H^0}$}
\newcommand{\cc}{cm\textsuperscript{-3}}

\newcommand{\te}{$T_e$}

\newcommand{\teoiiifir}{$T_{e\mathrm{,[O~III],FIR}}$}

\newcommand{\teniifir}{$T_{e\mathrm{,[N~II],FIR}}$}
\newcommand{\thi}{$T_\mathrm{H^0}$}

\newcommand{\msun}{$M_\odot$}

\newcommand{\um}{\micron}

\newcommand{\bsf}[1]{\textbf{\textsf{#1}}}
\newcommand{\zz}{\textit{z}}

\newcommand{\av}{$A_\mathrm{V}$}
\newcommand{\hardness}{$Q_1/Q_0$}
\newcommand{\lx}{$L_\mathrm{X}$}
\newcommand{\lbol}{$L_\mathrm{bol}$}

\newcommand{\cloudy}{\texttt{Cloudy}}

\defcitealias{paperi}{Paper I}
\newcommand{\ppi}{\citetalias{paperi}}

\newcommand{\ppiii}{Paper III}

\begin{document}

\title{Fine-structure Line Atlas for Multi-wavelength Extragalactic Study (FLAMES) II: \\Photoionization Model View of Ionized to Neutral Gas Emission}

\shorttitle{FLAMES II: A Cloudy View}
\shortauthors{Peng et al.}

\correspondingauthor{Bo Peng}
\email{bp392@cornell.edu}

\author[0000-0002-1605-0032]{Bo Peng}
\affiliation{Max-Planck-Institut für Astrophysik, Garching, D-85748, Germany} 
\email{bp392@cornell.edu}

\author[0000-0002-4444-8929]{Amit Vishwas}
\affiliation{Cornell Center for Astrophysics and Planetary Science, Cornell University, Ithaca, NY 14853, USA}
\email{vishwas@cornell.edu}

\author[0000-0003-1874-7498]{Cody Lamarche}
\affiliation{Department of Physics, Winona State University, Winona, MN 55987, USA}
\email{cody.lamarche@winona.edu}

\author{Gordon Stacey}
\affiliation{Department of Astronomy, Cornell University, Ithaca, NY 14853, USA}
\email{stacey@cornell.edu}

\author[0000-0002-1895-0528]{Catie Ball}
\affiliation{Department of Astronomy, Cornell University, Ithaca, NY 14853, USA}
\email{cjb356@cornell.edu}

\author[0000-0002-8513-2971]{Christopher Rooney}
\affiliation{National Institute of Standards and Technology, Boulder, CO 80305, USA}
\email{ctr44@cornell.edu}

\author{Thomas Nikola}
\affiliation{Cornell Center for Astrophysics and Planetary Science, Cornell University, Ithaca, NY 14853, USA}
\email{tn46@cornell.edu}

\author[0000-0001-6266-0213]{Carl Ferkinhoff}
\affiliation{Department of Physics, Winona State University, Winona, MN 55987, USA}
\email{cferkinhoff@winona.edu}

\begin{abstract}

Far-infrared (FIR) and mid-infrared (MIR) fine-structure lines (FSLs) provide key diagnostics of physical conditions in the interstellar medium (ISM). 
Building on empirical relations established in our previous work, we use photoionization models to systematically investigate the emission from both ionized and neutral gas phases in a coherent structure. 
By applying power-law fits to model parameters, we quantitatively capture how key FIR FSL ratios scale with physical properties such as density, radiation field strength and hardness, and elemental abundances. 
Our analysis confirms the primary dependencies behind most observed empirical trends and establishes certain FIR FSL ratios as tracers of physical parameters, while revealing that parameter marginalization---particularly in density and the O/H–
$U$–\hardness{} relation---plays a crucial role in shaping tight correlations seen in galaxies. 
We also identify persistent challenges, including degeneracies between ionization parameter and radiation field hardness, uncertainties in neutral gas density, and difficulties in modeling dusty \hii{} regions. 
We outline the fundamental observational and theoretical limitations of current FIR FSL diagnostics, and highlight prospects for advancing the field through comprehensive, multi-wavelength studies of diverse galaxy populations. 

\end{abstract}

\keywords{Fine-structure lines; Photoionization model; Interstellar medium; Chemical abundance; photodissociation regions}

\section{Introduction}
\label{sec:model_intro}

Far-infrared (FIR) and mid-infrared (MIR) fine-structure lines (FSLs) serve as powerful diagnostics for studying galaxy evolution. 
These lines contain key information about the physical conditions of the interstellar medium (ISM), including elemental abundances, the properties of radiation fields, and the density and structure of gas in star-forming regions. 
In a previous paper in this series \citep[][hereafter \ppi{}]{paperi}, we established empirical relationships among observed line ratios, focusing on the tracers of density, radiation field, abundance, and electron temperature. 
Nonetheless, the complex interdependence of these various properties renders the interpretation of FIR FSLs a non-trivial task. 

Theoretical studies of fine-structure lines have traditionally been divided into two separate regimes: ionized gas emissions are modeled using photoionization grids, while neutral gas lines are addressed within the photodissociation region (PDR) framework. 
This dichotomy complicates efforts to synthesize the information from all FSLs and hinders a comprehensive understanding of ISM properties. 
Only a few studies \citep[e.g.,][]{abel05,C19} have attempted to model both ionized and neutral gas emissions simultaneously. 
However, these works are limited either by their purely theoretical approach without observational comparison or by their focus on a narrow subset of galaxies with restricted property ranges. 
Additionally, FIR lines have been studied separately from optical lines because of the little overlap in samples and the large dust attenuation in dusty galaxies. 

The rapidly growing number of FIR FSL observations---particularly of \cii{} and \oiii{88}---and the considerable diversity in line ratios and luminosities observed in high-redshift (high-\zz{}) galaxies underscore the need for a more robust, systematic interpretive framework. 
Moreover, the strong correlations identified between ionized and neutral gas lines and between FIR and optical line diagnostics in \ppi{} motivate us to bridge the traditional divide and seek underlying connections between these ISM phases. 
To this end, we employ photoionization models extended to high optical depths, providing a unified theoretical perspective on FIR FSLs.

In this paper, we present the photoionization grids and evaluate their ability to reproduce the main diagnostic trends established in \ppi{}. 
Section~\ref{sec:model_method} describes the setup of our photoionization models and provides a detailed example of model structure. 
In Section~\ref{sec:model_result}, we examine how key diagnostics scale with model input parameters and discuss the theoretical interpretations of these diagnostics. 
Section~\ref{sec:model_discussion} addresses the ISM properties inferred from line diagnostics, discusses the challenges in constructing realistic dusty photoionization models, and outlines the limitations and future prospects of FSL studies.

\section{Method}
\label{sec:model_method}

\subsection{Photoionization Grid Setup}
\label{sec:model_grid}

In this work, we use \cloudy{} C23.01 release \citep{cloudy23,cloudy23.01} to construct a comprehensive grid of photoionization models, with careful consideration of the setups. 
The geometry is set to a plane-parallel configuration, providing a more general framework than the spherical geometry typically used for compact \hii{} regions, and thereby mitigating concerns regarding the origin of gas emission discussed in \ppi{}.

The incident radiation field that powers the gas comprises the following components:

\begin{itemize}
    \item \textbf{Stellar Radiation:} The POPSTAR 2009 stellar population synthesis models\footnote{Available at \url{https://www.fractal-es.com/PopStar}} \citep{molla09} are adopted as the input stellar spectral energy distributions (SEDs), using a Chabrier initial mass function \citep{chabrier03}. The model grid can span metallicities from 0.005 to 2.5 times solar, stellar ages from 0.1 Myr to 13.8 Gyr, and upper stellar masses up to 100\,\msun{}. The incident flux is varied along with the hydrogen density (\hdens{}) to cover a range of ionization parameters ($U$), where $U$ is defined as the ratio of the ionizing photon density to the electron density (\edens{}).
    \item \textbf{Cosmic Microwave Background (CMB):} The CMB at \zz{}=0 is included as part of the incident field.
    \item \textbf{X-ray Emission:} Bremsstrahlung emission at $T = 5 \times 10^6\,\mathrm{K}$ is added to account for potential X-ray heating in molecular gas. The X-ray luminosity is set at \lx{}~=~10\textsuperscript{-5}\,\lbol{}, balancing the lower value for O-type stars \citep{sciortino90} and the higher value used in \citet{C19}. The impact of X-rays is minor in our models, which focus on neutral gas.
    \item \textbf{Cosmic Rays:} The standard Galactic cosmic ray background intensity is included \citep{indriolo07}. 
\end{itemize}

Elemental abundances for \oh{} and \no{} are varied independently in the ranges --4.2 to --3.0 and --1.8 to --0.3, respectively. 
This independent variation in N/O is motivated by the secondary origin of nitrogen \citep{edmunds78,henry00,pilyugin03} and the diagnostic importance of the \nii{} line in both optical and FIR regimes. 
For other elements, solar abundance ratios (He/H, C/O, Ne/O, Na/O, Mg/O, Si/O, S/O, Cl/O, Ar/O, Ca/O, Fe/O) from \citet{asplund09} and \citet{grevesse98} are adopted, based on the production channel of each element. 
The same O/H is applied to the stellar population synthesis model.

Dust is a critical component in both the absorption of ionizing photons and the heating of neutral/molecular gas \citep{tielens85}. 
We assume a dust-to-gas mass ratio (D/G) of 0.01 at solar metallicity \citep{draine07,leroy23}, and simply scale D/G directly with O/H. 
Polycyclic aromatic hydrocarbons (PAHs), which are especially important for photoelectric heating in neutral gas \citep{dhendecourt87}, are included with abundances $n_\mathrm{C_{PAH}}/n_\mathrm{C} = 3.5 \times 10^{-2}$ and $n_\mathrm{C_{PAH,cluster}}/n_\mathrm{C} = 2 \times 10^{-2}$ \citep{tielens08}. 
PAH abundances are suppressed in ionized and molecular gas phases according to observational constraints \citep{deharveng10}.

A line width of 3\,km\,s$^{-1}$ is adopted, typical for the micro-turbulence in ISM. 
Magnetic fields are neglected for simplicity, as they contribute only to pressure in \cloudy{}. 
Gas structure is modeled under constant pressure, a choice justified in this section and discussed further in Sec.~\ref{sec:model_challenge}. 
Each model is computed to a depth corresponding to either \av{}~=~100\,mag or \te{}~=~10\,K, whichever is reached first. 
This prevents the model from extending too deeply into the molecular zone, where convergence issues become more likely. 

The model grid varies in five dimensions:

\begin{itemize}
	\item log $U$~=~--4 to --1, at an interval of 0.5. 
	\item log \hdens{} = 1.0 through 4.0, at an interval of 0.5 dex. 
	\item \no{}~=~--1.8 through --0.3, at an interval of 0.3 dex. 
	\item \oh{}~=~--4.2 through --3.0, at an interval of 0.2 dex. 
	\item age = 0.6, 1, 2, 3, 4, 5, 6, 10, 20, 30 Myr. 
\end{itemize}

\begin{figure}[]
	\centering
	\includegraphics[width=\halfwdth]{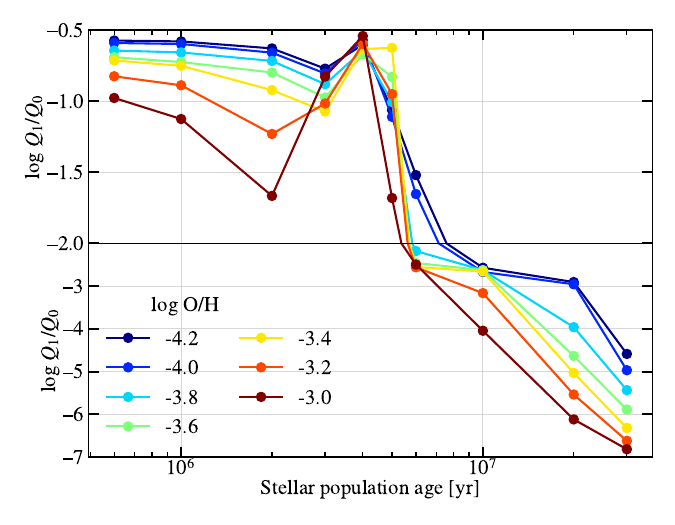}
	\caption[Stellar age--radiation hardness diagram.]{Stellar population age--radiation hardness diagram. The stellar population models of different metallicities are plotted in different colors. }
	\label{f:model_age-hardness}
\end{figure}

One key grid parameter, stellar age, does not directly affect ISM conditions but controls the spectral hardness of the ionizing field in a non-linear way, together with O/H. 
We define hardness (\hardness{}) as the luminosity ratio of helium-ionizing photons ($Q_1$) to total ionizing photons ($Q_0$). 
Fig.~\ref{f:model_age-hardness} illustrates the dependence of \hardness{} on stellar age for different metallicities. 
For young stellar populations ($\sim$Myr timescales), log (\hardness{}) typically ranges from --1.5 to --0.5. 
Given its direct influence on the ionized gas structure, we use \hardness{} instead of stellar age as the fifth grid dimension in our subsequent analysis.

Although models are computed over a broad range of stellar ages, only those with age $\leq$6\,Myr are used for analysis. 
This restriction is imposed because massive stars have evolved off the main sequence after $\sim$6\,Myr, resulting in SEDs with insufficient hardness to reproduce observable high-ionization lines such as \oiii{}, contrary to observations.

Dust is included in both neutral and ionized gas, as its presence in \hii{} regions is well-established \citep{inoue02} and is importance in the heating of neutral gas in PDRs \citep{tielens85}. 
However, strong dust absorption within the ionized region can induce unphysically high radiation pressure and density gradients near the ionization front (see Sec.~\ref{sec:model_dust} for more details). 
To avoid this, radiation pressure is neglected in our constant-pressure models.

To assess the impact of our modeling choices, we also compute and compare additional model grids: (1) spherical geometry; (2) plane-parallel geometry with radiation pressure included and dust suppressed in the ionized gas by scaling D/G with (1~--~\hp{}/H); and (3) constant D/G with radiation pressure included, in both spherical and plane-parallel geometries. 
Case (1) introduces more geometric variables but only minor differences and is thus not further considered. 
Case (2) also yields only small changes to the \hii{}--PDR structure. 
Case (3), which includes radiation pressure with constant D/G, results in significant model differences; these implications are discussed in detail in Sec.~\ref{sec:model_dust}.

\subsection{Model Structure and Output}
\label{sec:model_structure}

\begin{figure*}[]
	\centering
	\includegraphics[width=\textwidth]{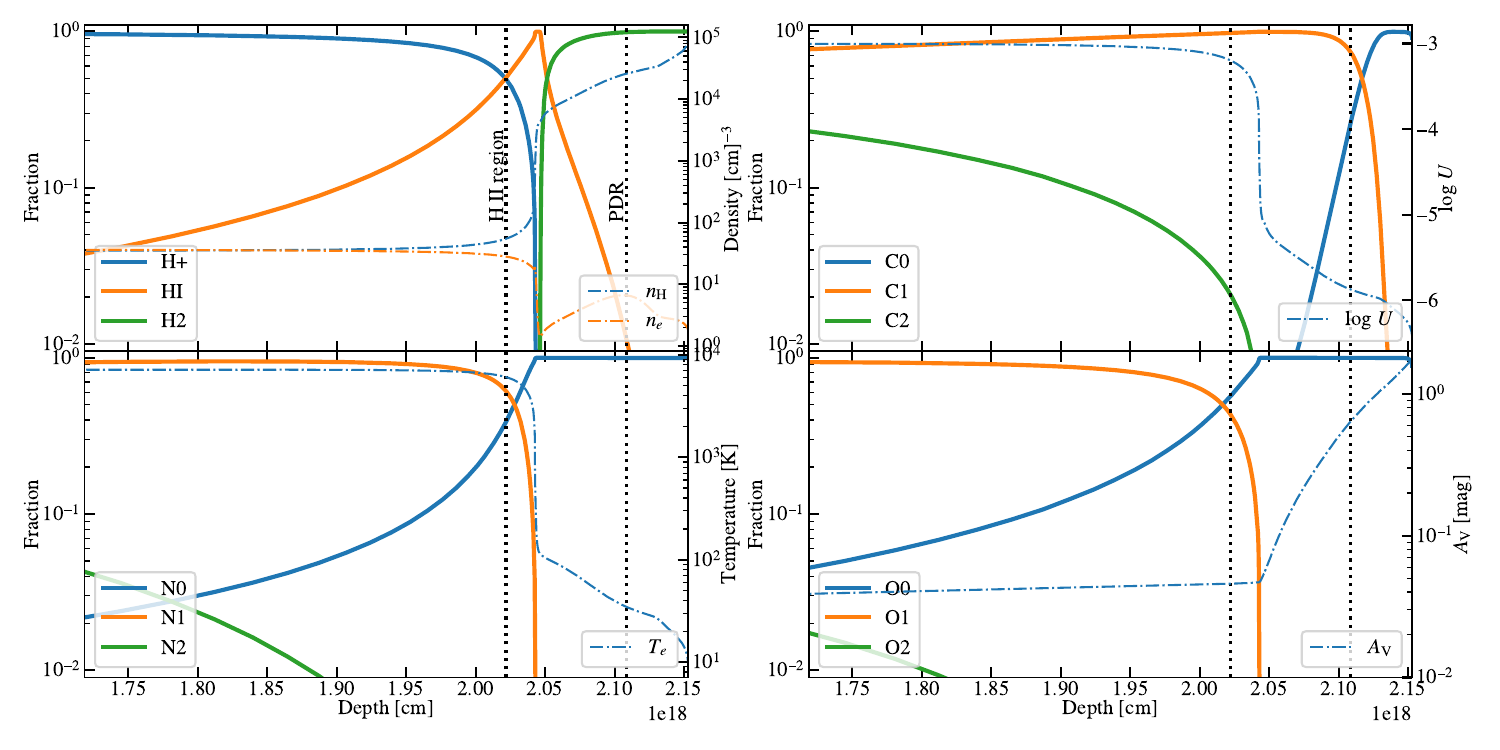}
	\caption[Example photoionization model structure.]{Example photoionization model structure of hydrogen (upper left), carbon (upper right), nitrogen (lower left), oxygen (lower right), as a function of depth. The y-axis on the right side shows the density (both total \hdens{} and electron \edens{}), ionization parameter (log $U$), electron temperature (\te{}), and extinction (\av{}), corresponding to the dash-dotted lines. Only the layer near the boundary of the \hii{} region is shown. }
	\label{f:model_structure}
\end{figure*}

As an illustrative example, we present the ionic structure of a representative photoionization model in Fig.~\ref{f:model_structure}. 
This model is set up with log $U$~=~--3.0, \hdens{} = 10\textsuperscript{1.5}\,\cc{}, \oh{}~=~--3.4, \no{}~=~--0.6, age~=~2 Myr. 

The boundary of the \hii{} region is defined as the position where the ionized hydrogen fraction (H\textsuperscript{+}/H) falls below 50\%. 
The PDR--molecular boundary is marked where the atomic hydrogen fraction (\hneut{}/H) drops below 1\%, indicating the dominance of molecular hydrogen. 
In this example, the distributions of \np{} and \op{} closely follow that of \hp{}, while \cp{} extends further, penetrating deep into the PDR and even into the molecular zone. 
Alongside the ionic fractions, we also plot the gas density, log $U$, electron temperature, and extinction throughout the model. 
Physical conditions remain relatively uniform within the \hii{} region, but change dramatically beyond the ionization front. 
The resulting ionized gas electron density and ionization parameter are consistent with the model’s initial setup.

\begin{figure*}[]
	\centering
	\includegraphics[width=\textwidth]{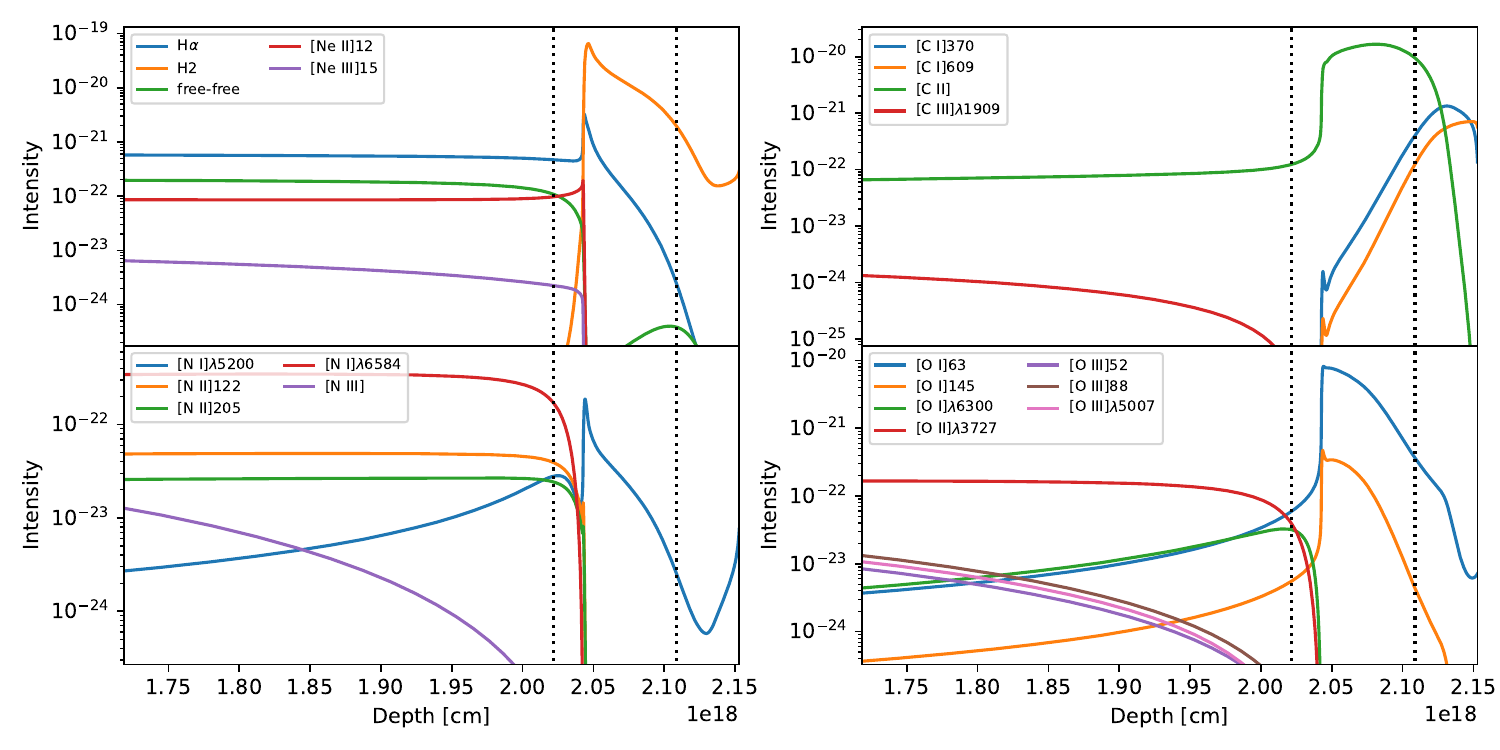}
	\caption[Example photoionization model emission intensity.]{Model output spectral line intensities near the \hii{} region boundary, for hydrogen \& neon (upper left), carbon (upper right), nitrogen (lower left), oxygen (lower right) spectral lines. The model output unit with arbitrary normalization is used.}
	\label{f:model_intensity}
\end{figure*} 

We also show the intensity profiles of metal lines from neutral to doubly ionized states, as well as several well-studied emission lines. 
It is important to note that our conservative definition of the \hii{} region allows some ionized gas emission to extend into the PDR; this does not impact our subsequent analysis. 
In this model, the emission peaks of the \cii{} and \oi{} lines are located within the PDR layer. 

A catalog of model output, including line fluxes and averaged physical properties, is provided in Table~\ref{tab:model_catalog} (see Appendix~\ref{sec:model_catalog}). 
Line fluxes are obtained by integrating the intensity over the full depth of the model, as is standard practice for \cloudy{} output \citep{vale16}. 
While emergent line fluxes are also computed, we generally do not use them except in cases where optical depth effects may be significant. 
This approach is motivated by the fact that models vary in column density and optical depth, leading to differences in dust attenuation and radiative transfer effects, which complicate the interpretation of emergent fluxes.

Model-averaged physical properties are calculated with appropriate weighting: without specification, quantities are weighted by volume; some quantities related to ionized gas (e.g., log $U$, \edens{}) are averaged using electron density weighting; while those relevant to neutral gas are averaged using hydrogen density weighting. 
Of the 20580 input models, 134 (0.65\%) failed to converge; the rest 20446 models are included in the catalog and used for subsequent analysis.

\section{Results}
\label{sec:model_result}

\subsection{Power Law Fit for Diagnostic Line Ratios}
\label{sec:model_fit}

A major challenge in interpreting the results of photoionization model grids lies in their high dimensionality. 
To accurately capture the influence of various physical parameters and ensure realistic modeling, it is necessary to allow as many parameters as possible to vary independently. 
However, this approach complicates the interpretation: observables are typically influenced by more than two parameters, rendering traditional grid diagrams insufficient to represent the full, multidimensional dependencies. 

Furthermore, the variation of input parameters in the models may not reflect the true range of variation observed in reality. 
For example, both the O/H and N/O abundance ratios exhibit significant variation in observations---well within the range spanned by our model grid---while other parameters, such as density, are inferred to vary less, as noted in \ppi{}. 
Consequently, large variations in observables caused by dominantly varying parameters (e.g., density) in the grid do not necessarily correspond to their real-world drivers. 
Conversely, parameters with weaker impact on the model, such as abundances, may be responsible for much of the observed diversity. 
Therefore, it is critical to quantitatively assess the dependence of each line ratio on all relevant parameters, not just those that cause the largest variation within the model grid. 

To achieve this, we perform multi-dimensional power-law fits for all line diagnostics examined in this work. 
Each ratio $R$ is fit using a least-squares approach to the following functional form:

\begin{equation}\begin{split}
	\log R =& x_0 + x_1 \log \left(\frac{U}{10^{-3}}\right) + x_2 \log \left(\frac{n_e}{\mathrm{cm^{-3}}}\right) + \\
	& x_3 \log \left(\frac{\mathrm{N/O}}{10^{-0.86}}\right) + x_4 \log \left(\frac{\mathrm{O/H}}{10^{-3.31}}\right) + \\
	& x_5 \log \left(\frac{Q_1/Q_0}{10^{-1}}\right)
\label{eq:fit}
\end{split}\end{equation}

Here, $U$ and \edens{} are the electron density---weighted values measured from the model outputs as described in Sec.~\ref{sec:model_structure}. 
The N/O and O/H terms use the input abundance ratios, normalized to their solar values. 
The radiation field hardness, quantified as \hardness{}, is used in place of stellar age because it directly characterizes the ionizing spectrum and thus the structure of the ionized gas; in contrast, stellar age is more relevant to galaxy evolution but does not directly determine line luminosities.

As shown in Fig.~\ref{f:model_age-hardness}, radiation field hardness is only weakly correlated with O/H, and this correlation is not monotonic---breaking down, for instance, between 3--5 Myr due to the Wolf–Rayet phase. 
Therefore, \hardness{} and O/H are treated as independent variables and both are included in the fitting procedure. 

\begin{rotatetable*}
\begin{deluxetable*}{lccccccc}
\centerwidetable
    \tablecolumns{8}
    \tablecaption{The power-law fitting results on line ratios and properties using the mathematical form Eq.~\ref{eq:fit}. The residual scatter in the logarithm space is reported in the last column. Diagnostics involving lines \oiii{} and \niii{} are performed on all relevant models as well as on the two parts of these models in the segmented fit. The conditions for the model selections are noted in the parenthesis, in the same order as the fitted values. \label{t:fit}}
    \tablehead{
	\colhead{(1)}	& \colhead{(2)}	& \colhead{(3)}	& \colhead{(4)}	& \colhead{(5)}	& \colhead{(6)}	& \colhead{(7)}	& \colhead{(8)}	\\
	\colhead{Quantity (fitting condition)}	& \colhead{$x_0$}	& \colhead{$x_1$}	& \colhead{$x_2$}	& \colhead{$x_3$}	& \colhead{$x_4$}	& \colhead{$x_5$}	& \colhead{$\Delta$}	\\
	& & \colhead{$\frac{U}{10^{-3}}$}	& 
    \colhead{$\frac{n_e}{50 \,\mathrm{cm^{-3}}}$}	& 
    \colhead{$\frac{\mathrm{N/O}}{10^{-0.86}}$} & 
    \colhead{$\frac{\mathrm{O/H}}{10^{-3.31}}$}	& 
    \colhead{$\frac{Q_1/Q_0}{0.10}$}	& \colhead{dex} 	}
    \startdata
    \cutinhead{Density}
	\hdens{} [\cc]	& 3.88	& -0.06	&  0.84	& 	& -0.51	& 0.10	& 0.14		\\
	\oi{145}/\cii{}		& -1.72	& 0.24	&  0.77	& 	& 0.10	& -0.13	& 0.23	\\
	\oi{63}/\oi{145}	& 1.36	& -0.05	&  0.04	& 	& -0.14	& 0.05	& 0.08	\\
    \cutinhead{Radiation field}
	\neiii{15}/\neii{12} & -1.15	& 0.89	& 	&  	& -0.52 & 2.72 	& 0.20	\\
	\oiii{88}/\cii{} (all,\textgreater 1,\textless 1)	& -1.11,-0.12,-1.02	& 1.19,0.62,1.78	& 0.08,0.13,0.04	&  	& 0.49,0.39,0.59 & 1.18,0.47,1.85	& 0.38.0.16,0.24	\\
	\niii{}/\nii{122} (all,\textgreater 5,\textless 5)	& 0.03,0.60,0.14	& 1.05,0.70,1.64	& 0.15,0.12,0.19	&  	& -0.21,-0.16,-0.27 & 0.76,0.49,1.07	& 0.27,0.15,0.09	\\
	\niii{}/\nii{205} (all,\textgreater 10,\textless 10)& 0.37,0.86,0.43	& 1.04,0.77,1.59	& 0.49,0.43,0.55	&  	& -0.19,-0.15,-0.25 & 0.75,0.52,1.03	& 0.30,0.21,0.13	\\
	\oiii{88}/\oi{145} (all,\textgreater 3,\textless 3)	& 0.67,1.13,0.90	& 0.93,0.61,1.51	& -0.71,-0.57,-0.88	&	& 0.37,0.33,0.42	& 1.24,0.75,1.86	& 0.35,0.22,0.20 \\
	\oiii{88}/\nii{122} (all,\textgreater 25,\textless 25) & 0.41,1.00,0.40	& 1.14,0.81,1.57	& -0.05,-0.06,-0.03	& -0.99,-0.93,-1.01	& -0.26,-0.27,-0.23	& 1.28,0.83,1.77 	& 0.28,0.15,0.17	\\
	\oiii{88}/\nii{205} (all,\textgreater 50,\textless 50) & 0.75,1.30,0.73	& 1.14,0.84,1.56	& 0.30,0.26,0.33	& -0.99,-0.91,-0.98	& -0.24,-0.24,-0.22	& 1.28,0.89,1.72 	& 0.31,0.22,0.21	\\
	\oiiio{5007}/\siio{6716,6731} (all,\textgreater 3,\textless 3)	& -0.69,0.12,-0.65	& 1.28,0.83,1.98	& 0.19,0.17,0.21	& 	& -0.65,-0.58,-0.74	& 1.16,0.56,1.88	& 0.35,0.14,0.20	\\
	\oiiio{5007}/\oio{6300} (all,\textgreater 100,\textless 100)	& 0.86,1.88,1.04	& 1.25,0.62,2.13	& 	& 	& -0.42,-0.46,-0.39	& 0.35,-0.22,0.97	& 0.41,0.10,0.17	\\
	\oiiio{5007}/\niio{6584} (all,\textgreater 10,\textless 10)	& -0.20,0.36,-0.21	& 1.23,0.92,1.68	& 0.05,0.03,0.07	& -1.02,-0.96,-1.06	& -0.62,-0.58,-0.66	& 1.36,0.94,1.83	& 0.27,0.13,0.18 \\
    \cutinhead{Abundance}
	\niii{}/\oiii{88}	& -0.38	& -0.09	& 0.20	& 1.00	& 0.05	& -0.52	& 0.12	\\
	\niii{}/\oiii{52,88}& -0.66	& -0.10	& -0.06	& 0.99	& 0.04	& -0.53	& 0.11	\\
	\nii{122}/\cii{}	& -1.58	& 0.13	& 0.14	& 0.98	& 0.68	& -0.12	& 0.27	\\
	\nii{205}/\cii{}	& -1.92	& 0.13	& -0.20	& 0.97	& 0.65	& -0.13	& 0.26	\\
    \cutinhead{Temperature}
	\te{} [K]		& 3.81	& 0.06	&   	&   & -0.29 & 0.08 	& 0.05		\\
	$T_{e\mathrm{,PDR}}$ [K]		& 2.42	& 	& 0.07 	&   & 0.20 & 0.10	& 0.21		\\
\enddata
\end{deluxetable*}
\end{rotatetable*}

\begin{figure}[]
	\centering
	\includegraphics[width=\halfwdth]{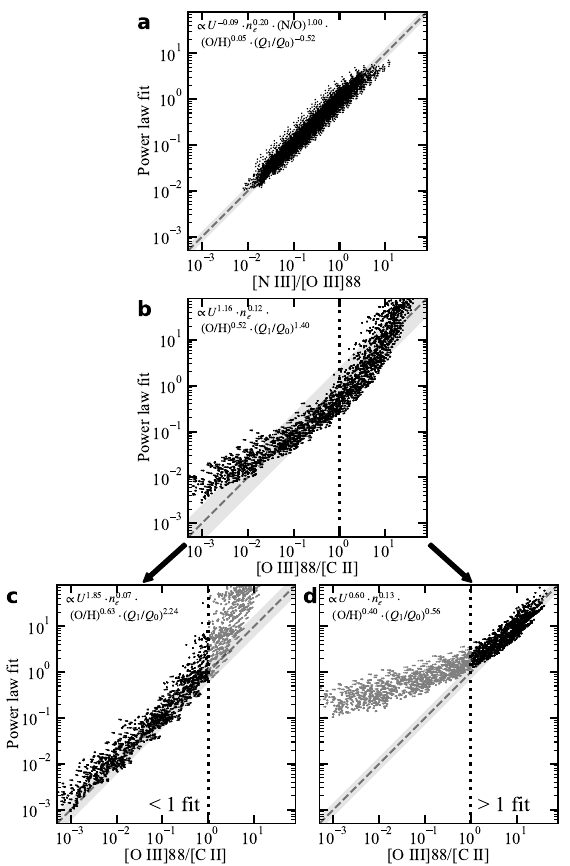}
	\caption[Examples of power-law fit.]{Examples of power-law fit to the model output of \bsf{a}, \cii{}/\ha{}; \bsf{b}, all models of \oiii{88}/\cii{}; \bsf{c}, \oiii{88}/\cii{} with the line ratio \textless 1; \bsf{d}, \oiii{88}/\cii{} with the line ratio \textgreater 1. In each panel, the model output quantity is plotted on x-axis, and the fitted relation is plotted as y. The thick gray dashed line denotes the 1:1 relation, and the shade depicts the residual scatter. In the segmented fitting shown in (\bsf{c}) and (\bsf{d}), the data used in the fitting are plotted in black, and those excluded by the selection in gray. }
	\label{f:model_fit}
\end{figure}

All power-law fitting results are summarized in Table~\ref{t:fit}, with example fits for \niii{}/\oiii{88} and \oiii{88}/\cii{} shown in Fig.~\ref{f:model_fit}. 
In reporting the fits, we omit any power-law indices smaller than 0.05, as these terms contribute negligibly compared to other parameters. 
The only exceptions are fits for \oi{63/145} and electron temperature, where the overall variation is so low that even the largest indices do not exceed 0.3.

A special scenario arises in fits involving both high-ionization lines (e.g., \oiii{} or \niii{}) and low-ionization lines, where the relationship exhibits a break, resulting in a segmented---or broken---power law. 
For these cases, Table~\ref{t:fit} includes both the fit over the entire data range and separate fits above and below the break point. 
The implications of such breaks are further discussed in Sec.~\ref{sec:model_radiation}. 

There are also cases where a simple power-law fit cannot adequately capture the model output, with residual scatter exceeding 0.5 dex. 
This is particularly true for line ratios that intrinsically depend on exponential functions of electron temperature, such as optical forbidden line-to-\ha{} or certain FIR FSL ratios. 
While \te{} itself may be reasonably fit by a power law (see Table~\ref{t:fit}), any ratio involving $e^{T_e}$ introduces non-linearity that invalidates the power-law approximation. 
Fortunately, most FSL ratios analyzed in this work are relatively insensitive to \te{}, allowing for straightforward interpretation using our fitting approach.
Additional remarks on the limitations or particularities of specific fits are provided in the relevant sections throughout the paper.

\subsection{Density}
\label{sec:model_ne}

\begin{figure}[]
	\centering
	\includegraphics[width=\halfwdth]{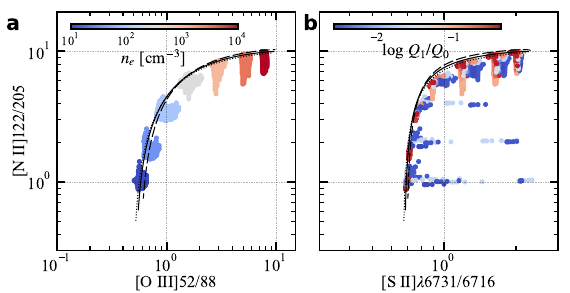}
	\caption[\nii{122/205} vs. high density tracers. ]{Comparison of model \nii{122/205} to \bsf{a}, \oiii{52/88}; or \bsf{b}, \siio{6731/6716}. Data points are color-coded according to the color bars displayed. The solid, dashed, and dotted thin lines are the theoretical emissivity ratio calculated using \te{} = 1E4, 2E4, 5E3 K, the same as those in fig.~17 in \ppi{}.}
	\label{f:model_n_NII-n_ion}
\end{figure}

Following the approach of \ppi{}, we begin by examining density diagnostics. 
For this analysis, high-density tracers are excluded, as observed values cluster around their low-density limits. 
Fig.~\ref{f:model_n_NII-n_ion} presents a comparison between the low-density electron density tracer \nii{122/205} and its high-density counterparts, \oiii{52/88} and \siio{6731/6716}, using the same coordinate ranges and theoretical emissivity ratio curves as fig.~13 (\bsf{c}) and (\bsf{d}) in \ppi{} to facilitate direct comparison.

The models predict that regions traced by \opp{} are slightly denser than those traced by \np{}, with a similar trend observed for \spion{}. 
However, the degree of density stratification in the models is insufficient to account for the scatter and trends seen in the observational data. 
For the [S~{\sc ii}]-\nii{} electron density comparison, the model does produce some points with significant stratification, but these are primarily associated with lower log (\hardness{})~\textless~--2, indicative of older stellar populations with a reduced output of hard ionizing photons. 
Overall, the models do not reproduce the full scatter and deviation observed in the data. 
This discrepancy may reflect either limitations in the physical structures represented by the models, or significant observational uncertainties not captured by the reported error bars---the latter interpretation is also supported by the scatter seen in high-density tracer comparisons.

\begin{figure}[]
	\centering
	\includegraphics[width=\halfwdth]{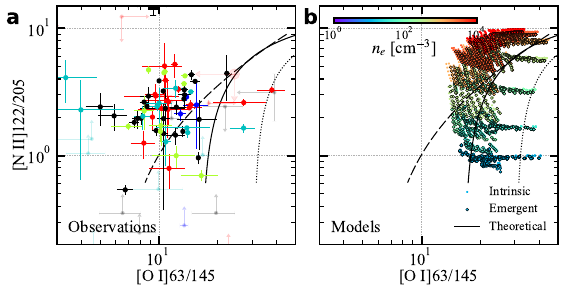}
	\caption[\nii{122/205} vs. \oi{63/145}.]{\nii{122/205} to \oi{63/145} of \bsf{a}, observations; and \bsf{b}, models. The color classification by galaxy type in the left panel is the same as that in fig.~1 in \ppi{}. For the model points, both the intrinsic flux by integrating intensity and emergent flux by considering radiative transfer effect ratios are plotted as color disk without edge and diamond with black edge. The color-coding of model points is based on \edens{} corresponding to the color bar. The solid, dashed, and dotted thin lines are the same emissivity ratio curves as those in \ppi{} fig.~13 (\bsf{e}). The observational data in (\bsf{a}) use the FLAMES-low and high table in \ppi{}, the plotted data points are sourced from \input{n_NII-n_OI_low}, \input{n_NII-n_OI_high}, and Peng et al. in prep.}
	\label{f:model_n_NII-n_OI}
\end{figure}

Regarding \oi{63/145}, Fig.~\ref{f:model_n_NII-n_OI} displays the comparison to \nii{122/205}. 
The model distributions are broadly consistent with theoretical predictions for the emissivity ratio under pressure equilibrium and \thi{}~=~150\,K (solid line), as also shown in fig.~13 (\bsf{e}) of \ppi{}. 
However, the model predicts systematically higher \oi{63/145} ratios than observed values, which cluster around 10. 
To assess the impact of \oi{63} self-absorption, we also plot the emergent line ratio from \cloudy{} output (diamonds), which shows only a slight reduction relative to the intrinsic ratio, indicating minimal self-absorption in the model.

\begin{figure}[]
	\centering
	\includegraphics[width=\halfwdth]{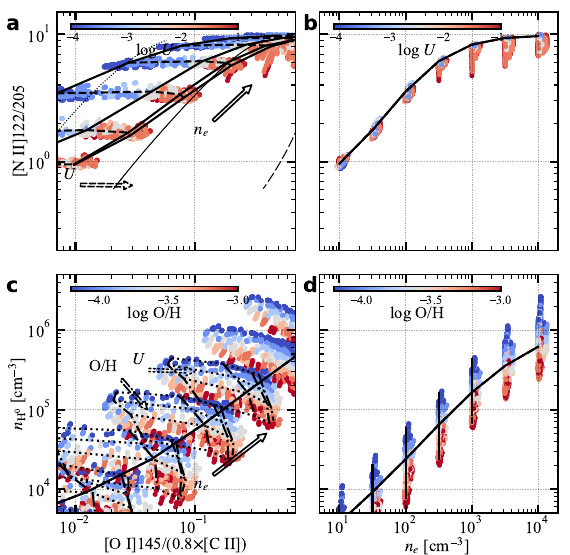}
	\caption[\nii{122/205} vs. \oi{145}/\cii{}. ]{Model interpretation of \nii{122/205} vs. \oi{145}/\cii{} comparison. \bsf{a}, shows the model output line ratios color-coded by log $U$, and the 2-d grid manifests the effect of \edens{} increasing along the thick solid lines, and $U$ increasing along the dashed lines, across the whole range of the two parameters with all other parameters fixed. The solid, dashed, dotted thin lines are the same emissivity curves as those in \ppi{} fig.~13 (\bsf{f}). \bsf{b}, shows \nii{122/205} varying by \edens{}. \bsf{c}, shows the relation between atomic Hydrogen density in neutral gas and \oi{145}/\cii{}, the solid line shows the effect of increasing \edens{}, and subgrid of O/H (along dash-dotted lines) and $U$ (dotted lines) are shown at each grid value of \edens{}. \bsf{d}, plots the atomic density \hdens{} against electron density \edens{}. The diagonal solid line denotes varying \edens{}, while the vertical lines correspond to the same subgrid as that in the lower left panel, but the deviation is dominated by the effect of O/H in this case. }
	\label{f:model_n_NII-n_neut_model}
\end{figure}

The comparison of \nii{122/205} to \oi{145}/\cii{} is shown in Fig.~\ref{f:model_n_NII-n_neut_model}. 
The \oi{145}/\cii{} ratio is primarily sensitive to \edens{}, but also exhibits dependence on $U$ and O/H. 
At fixed electron density, \oi{145}/\cii{} increases with $U$ for log $U$~$\lesssim$~--3, but this trend weakens for higher ionization parameters. 
The ratio also shows a non-negligible dependence on O/H (power-law index $\sim$0.08). 
(\bsf{c}) reveals that \oi{145}/\cii{} is influenced by not only atomic hydrogen density, but also a combination of $U$ and O/H that affect neutral gas temperature.

When comparing the photoionization models to the theoretical emissivity ratios under pressure equilibrium (thin solid line in (\bsf{a})), the models tend to predict lower \oi{145}/\cii{} values at low electron densities. 
This suggests that the neutral gas in the models is less dense than simple pressure equilibrium, likely because the assumption of constant temperature contrast does not hold, and the neutral gas cools more rapidly than the ionized gas as density decreases. 

Observational data (see fig.~13 in \ppi{}) cluster at \nii{122/205} $\sim$1.5-4 and \oi{145}/(0.8$\times$\cii{}) $\sim$0.03--0.2, corresponding to electron densities of $\sim$20--100\,\cc{} and atomic hydrogen densities of 10\textsuperscript{4}--10\textsuperscript{5}\,\cc{} based on the model grid. 
The data distribution also suggests that most galaxies have log $U$~\textgreater~--3, though higher values are not well constrained. 
As the observations are scattered around the theoretical emissivity predictions, the models may underestimate the neutral gas density---a possibility discussed further in Sec.~\ref{sec:model_dust}.

\subsection{Radiation Field}
\label{sec:model_radiation}

As demonstrated in \ppi{}, high-ionization state ions provide powerful diagnostics of the ionizing radiation field. 
The relative proportions of high- and low-ionization states are primarily set by the competition between photoionization and recombination, thus $U$ is defined as the ratio of the ionizing photon density to the electron density. 
Photoionization also requires photons of sufficient energy to ionize each species, so the presence of higher-state ions is a sensitive tracer of a harder radiation field, i.e., larger values of \hardness{}. 

Most high-to-low ionization line ratios depend on both $U$ and \hardness{}. 
For ions with ionization potentials similar to helium, ratios such as \oiii{}/\cii{}, \oiii{}/\nii{} (both FIR and optical), and \niii{}/\nii{} show power-law indices for $U$ and \hardness{} of comparable magnitude in our fitting results. 
These ratios are thus controlled not by $U$ or \hardness{} alone, but by their product, $U_1 = U \times (Q_1/Q_0)$, which can be interpreted as a helium ionization parameter. 
For ions with even higher ionization potentials, the dependence on $U$ remains close to linear, while the dependence on \hardness{} becomes much steeper. 
For example, \neiii{}/\neii{} scales as $U \times (Q_1/Q_0)^3$. 

The behavior of the \oiii{} and \niii{} ratios to lower ions is more complex. 
At large high-to-low ionization ratios (e.g., large \oiii{}/\cii{} or \niii{}/\nii{}), the dependence on $U$ and \hardness{} is moderate (power-law indices $\sim$0.6--0.7), but as the ratio decreases, the power-law indices increase by more than one dex. 
This trend is evident in Fig.~\ref{f:model_fit}, where a break occurs at \oiii{88}/\cii{}~$\sim$~1, and segmented power-law fits above and below this threshold provide better agreement than a global fit. 
All \oiii{} and \niii{} fits in Table~\ref{t:fit} are reported for both global and segmented fits. 
This break is the result of a shift in the dominant ionization stage of oxygen and nitrogen in the ionized gas, from \opp{} and \npp{} to \op{} and \np{}.

\begin{figure}[]
	\centering
	\includegraphics[width=\halfwdth]{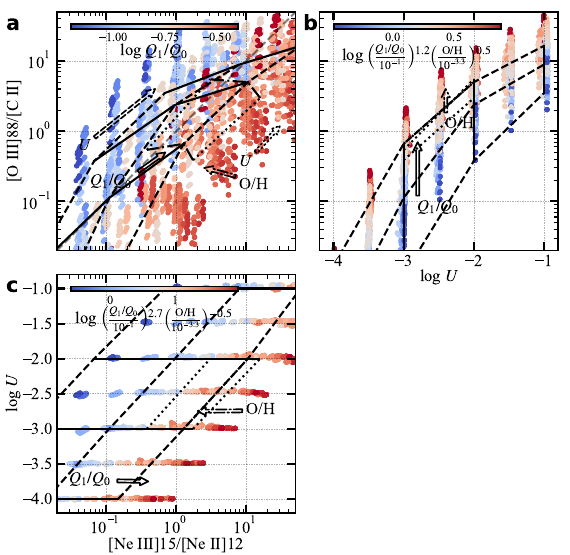}
	\caption[\neiii{15}/\neii{12} vs. \oiii{88}/\cii{}. ]{Model interpretation of \neiii{15}/\neii{12}--\oiii{88}/\cii{} diagnostics. \bsf{a}, the model data points comparing the two line ratios are plotted and color-coded by \hardness{}. The 2-d grid shows the effect of increasing \hardness{} (log (\hardness{})~=~--1.4, --1.0, to --0.6) along the solid lines, and increasing $U$ along the dashed lines. At log $U$~=~--3 and --2, a subgrid of O/H (dash-dotted lines at \oh{}~=~--3.8, --3.4, --3.0) and $U$ (dotted line) is also plotted. The diagonal gray dashed thick line is a visual aid for linearity. \bsf{b}, illustrates the dependence of \oiii{88}/\cii{} on log $U$, and the additional impact by \hardness{} and O/H. \bsf{c}, depicts the similar relation of \neiii{15}/\neii{12}. Note that the color-coding of the $U$ dependence is the power-law fit excluding the $U$ term. }
	\label{f:model_OIII_CII-NeIII_NeII}
\end{figure}

Although $U$ and \hardness{} are degenerate in many high-ionization diagnostics as $U_1$ is effectively in control, the steep dependence of \neiii{15}/\neii{12} on \hardness{} helps break this degeneracy. 
In Fig.~\ref{f:model_OIII_CII-NeIII_NeII}, the model grid forms a 2D structure where \oiii{88}/\cii{} increases mainly with $U$, while \neiii{15}/\neii{12} is particularly sensitive to \hardness{}. 
This distinction is most pronounced at high \oiii{88}/\cii{}, where the diagnostic power of \oiii{88}/\cii{} for \hardness{} drops due to the aforementioned break in scaling. 
In addition, both ratios show a secondary dependence on O/H, and the model grids in Fig.~\ref{f:model_OIII_CII-NeIII_NeII} includes subgrids to illustrate this effect.

Compared to the observed line ratio distributions (fig.~16 (\bsf{a}) in \ppi{}), most local U/LIRGs have log $U$ between --3 and --2.5 and log  (\hardness{}) around --1, while dwarf galaxies extend to higher log $U$ ($\sim$--2) and harder radiation (log (\hardness{})~$\sim$~--0.6). 
Due to their systematically lower O/H, actual $U$ values in dwarfs may be even higher. 
Overall, the large spread observed in \oiii{88}/\cii{} and related diagnostics primarily reflects variations in $U$ across galaxies, with additional effects of \hardness{} and O/H. 

\begin{figure}[]
	\centering
	\includegraphics[width=\halfwdth]{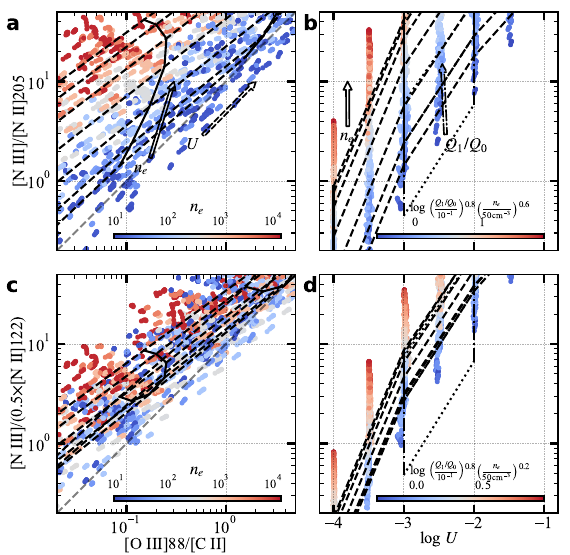}
	\caption[\oiii{88}/\cii{} vs. \niii{}/\nii{} doublets. ]{Model view of the comparison of \oiii{88}/\cii{}--\niii{}/\nii{}. Both \nii{205} (top row) and \nii{122} (bottom row) are shown. The right column shows the \niii{}/\nii{} dependence on log $U$. In left column, the grid of \edens{} (solid lines) and $U$ (dotted lines) is shown. The diagonal gray dashed thick line is a visual aid for linearity. In right column, in addition to the main \edens{}--$U$ grid, an O/H (dash-dotted lines)--$U$ (dotted lines) subgrid is also plotted. }
	\label{f:model_OIII_CII-NIII_NII}
\end{figure}

We compare the two main radiation field diagnostics in Fig.~\ref{f:model_OIII_CII-NIII_NII}. 
Both \oiii{}/\cii{} and \niii{}/\nii{} are primarily sensitive to $U_1$, with modest differences in their \hardness{} dependence. 
The electron density also plays a role, especially for \niii{}/\nii{205} due to the low critical density of \nii{205}. 
The vertical offset between \oiii{}/\cii{} and \niii{}/\nii{205} serves as a probe of electron density, whereas \niii{}/\nii{122} is preferable for pure radiation field diagnostics. 
Observationally (see fig.~17 (\bsf{b}) in \ppi{}), the line ratios are scattered around the diagonal, with offsets larger than predicted even for the lowest model electron densities (\edens{}~$\sim$~10\,\cc{}). 
This may be a result of AGN contamination in the \oiii{} emission, as AGN hosts appear to the right of the diagonal while U/LIRGs group to the left, in better agreement with model predictions.

\begin{figure*}[]
	\centering
	\includegraphics[width=0.200\textwidth]{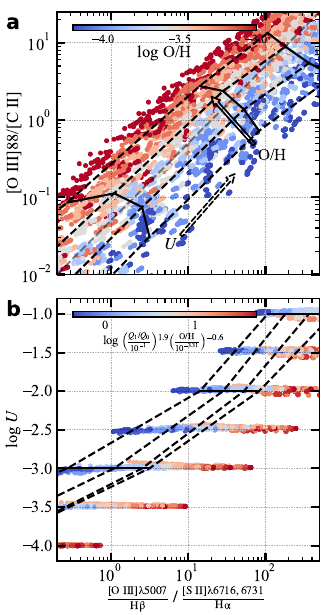}
	\includegraphics[width=0.377\textwidth]{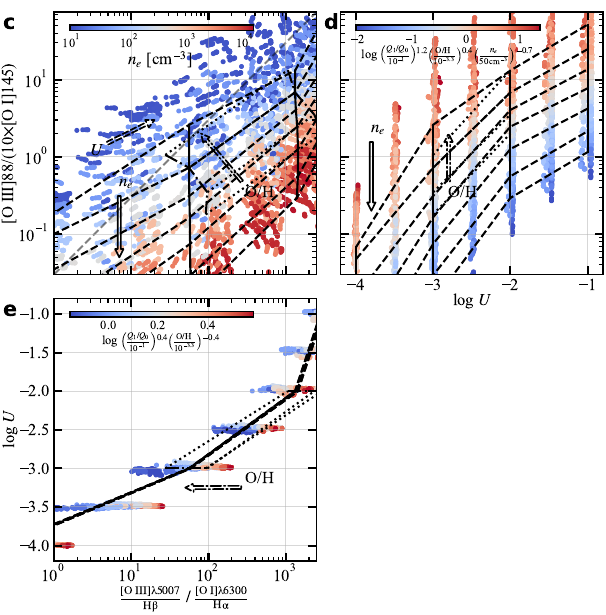}
	\includegraphics[width=0.382\textwidth]{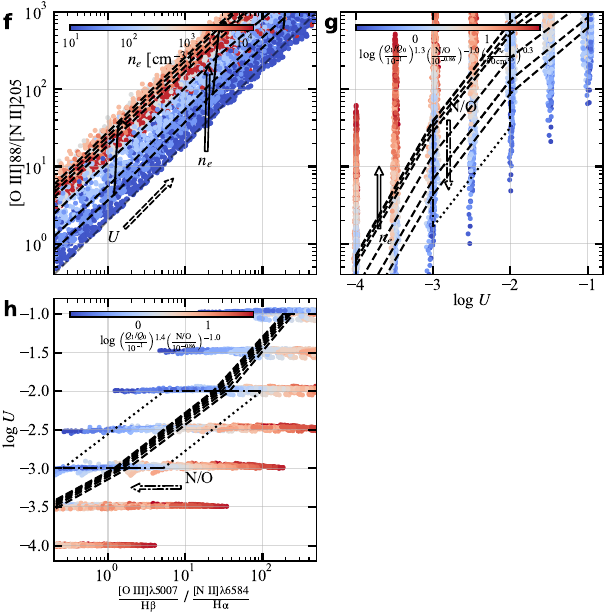}
	\caption[Optical vs. FIR radiation field diagnostics. ]{Radiation field diagnostics comparison between optical and FIR line ratios. \bsf{a}-\bsf{b} (left), \oiiio{5007}/\siio{6717,6731} vs. \oiii{88}/\cii{}; bsf{c}-\bsf{e} (middel), \oiiio{5007}/\oio{6300} vs. \oiii{88}/\oi{145}; bsf{f}-\bsf{h} (right),  \oiiio{5007}/\niio{6584} vs. \oiii{88}/\nii{205}. Subpanels illustrating the effect of log $U$ are shown adjunct to the main panels, similar to the format of Fig.~\ref{f:model_OIII_CII-NeIII_NeII}. In the left panels, a grid of O/H (solid lines from --4.2 to --3.0)--$U$ (dashed lines) is shown. In the middle panels, the grid of \edens(solid lines)--$U$(dashed lines) and the subgrid of O/H (dash-dotted lines)--$U$ (dotted lines) are shown. In the right panels, the grid of \edens(solid lines)--$U$(dashed lines) and the subgrid of N/O (dash-dotted lines)--$U$ (dotted lines) are shown. The gray dashed thick line in each main panel denotes the linearity as a visual aid.}
	\label{f:model_U-U}
\end{figure*}

In addition to FIR diagnostics, optical \oiii{} lines serve as effective tracers of the radiation field. 
Fig.~\ref{f:model_U-U} compares FIR and optical diagnostics in the same parameter space as fig.~28 in \ppi{}. 
All ratios are primarily driven by changes in $U_1$, with additional dependencies on \edens{} (for diagnostics involving \oi{145} or \nii{205}) and secondary correlations with O/H (notably in \oiiio{5007}/\siio{6716,6731}). 
Most ratios have similar power-law indices for $U$ and \hardness{}, though \oiiio{5007}/\oio{6300} shows a notably weaker dependence on \hardness{}, making it a useful tool for breaking $U$–\hardness{} degeneracy at low \oiii{88}/\cii{}. 
The observed distribution of \oiii{}/\oi{} along the diagonal supports the conclusion that $U$ is the primary driver of \oiii{} line variations, with \hardness{} playing a lesser role.

\subsection{Abundance}
\label{sec:model_metallicity}

\begin{figure}[]
	\centering
	\includegraphics[width=\halfwdth]{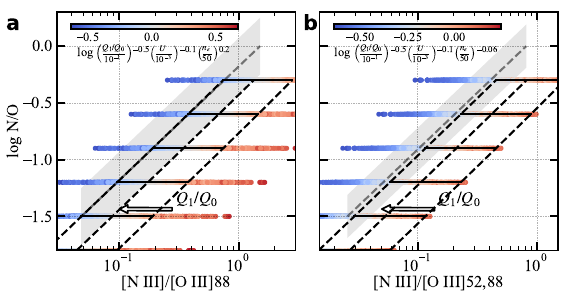}
	\caption[\niii{}/\oiii{} diagnostics. ]{N/O diagnostics using \bsf{a}, \niii{}/\oiii{88}; and \bsf{b}, \niii{}/\oiii{52,88}. The gray and thick dashed lines and the shade represent the fit on the observational data in \ppi{} fig.~19. A grid of \hardness{} (solid lines, --0.6, --1.0, --1.4)--N/O is also plotted. }
	\label{f:model_NIII_OIII}
\end{figure}

The \niii{}/\oiii{} ratio is a robust diagnostic for the N/O abundance ratio, owing to the similar ionization potentials of the \op{} and \np{} ions to \opp{} and \npp{}. 
Our fitting results confirm that \niii{}/\oiii{} varies linearly with N/O, with \hardness{} identified as the next most significant factor. 
As shown in the N/O--\hardness{} grid in Fig.~\ref{f:model_NIII_OIII}, the influence of radiation hardness on this diagnostic is non-negligible, consistent with the finding in \citet{P21}. 

Although the model outputs for \niii{}/\oiii{} span a wide range at fixed N/O, the observed relation (shaded region in Fig.~\ref{f:model_NIII_OIII}) exhibits much less scatter, typically within 0.2 dex. 
This suggests that, in real galaxies, either the other relevant parameters (e.g., \edens) vary over a much narrower range than in the model grid, or their effects cancel out due to correlations among galaxy properties. 
In Fig.~\ref{f:model_NIII_OIII}, both \niii{}/\oiii{88} and \niii{}/(\oiii{52}+\oiii{88}) are plotted. 
The density dependence of these ratios is 0.17 and --0.08 dex, respectively, indicating only a slight improvement for the latter, as also noted in \ppi{}. 

\begin{figure}[]
	\centering
	\includegraphics[width=\halfwdth]{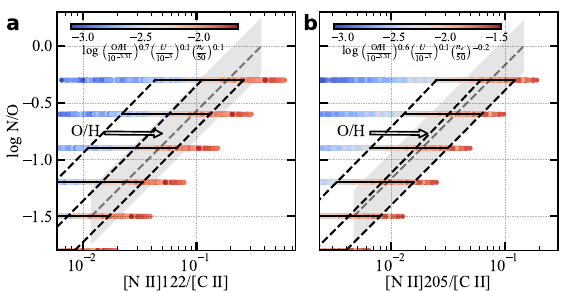}
	\caption[\nii{}/\cii{} diagnostics. ]{N/O diagnostics using \nii{122}/\cii{} (left) and \niii{205}/\cii{} (right). The gray and thick dashed lines and the shade represent the fit on the observational data in \ppi{} fig.~20. A grid of O/H (solid lines, --3.8, --3.4, --3.0)--N/O is also plotted. }
	\label{f:model_NII_CII}
\end{figure}

The \nii{}/\cii{} ratio is also a useful tracer of N/O, and our fits confirm that it is linearly related to N/O. 
However, this diagnostic is more strongly affected by other parameters, particularly O/H, which enters with a power-law index of --0.64. 
Additional, though less significant, dependencies on $U$ and \hardness{} are also present. 
Both \nii{122}/\cii{} and \nii{205}/\cii{} are subject to density effects, although the directions of dependence differ between the two ratios when considering total \cii{} emission. 
As with \niii{}/\oiii{}, the observed relation between the line ratio and N/O displays much less scatter than the models. 
To reconcile this, the negative dependence on O/H ($\mathrm{O/H}^{-0.64}$) need to be largely compensated for by intrinsic correlations among O/H, $U$, and \hardness{} in observed galaxies.

\subsection{Electron Temperature}
\label{sec:model_te}

\begin{figure}[]
	\centering
	\includegraphics[width=\halfwdth]{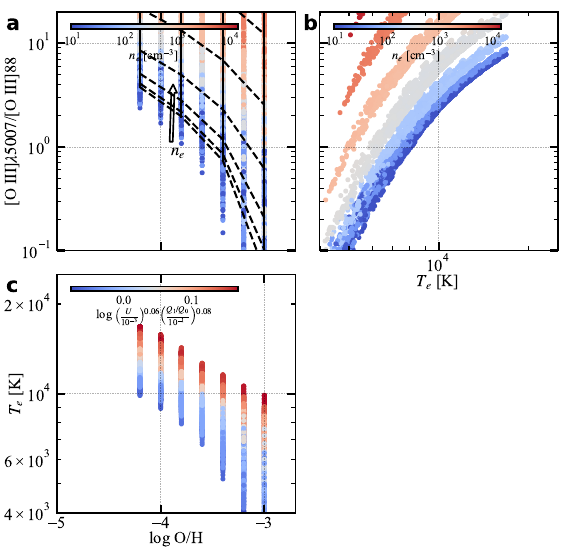}
	\caption[\teoiiifir{} diagnostics. ]{\oiiio{5007}/\oiii{88}--\oh{}, with the grid \edens{} (solid lines)--O/H (dashed lines) overplotted on model points in \bsf{a}. }
	\label{f:model_OIII_te}
\end{figure}

Fig.~\ref{f:model_OIII_te} shows the model predictions for the diagnostic ratio \oiiio{5007}/\oiii{88}. 
The \oiiio{5007}/\oiii{88} versus \te{} diagram exhibits a very tight correlation, with the separation between model points solely determined by electron density (\edens{}). 
This is expected as both \te{} and \edens{} have a direct impact on the ionized line ratios. 
In our models, \te{} is found to depend mainly on O/H abundance and also on $U_1$, reflecting the key role of these parameters in setting the heating–cooling balance in the ionized gas.

The observed anti-correlation between \oiiio{5007}/\oiii{88} and O/H, as seen in fig.~23 of \ppi{}, is well reproduced by our model output. 
The vertical offset in the \oiiio{5007}/\oiii{88}–O/H relation provides a sensitive measure of \edens{}, and the data suggest \edens{}~\textless~100\,\cc{}---in agreement with constraints from other density tracers. 

\begin{figure}[]
	\centering
	\includegraphics[width=\halfwdth]{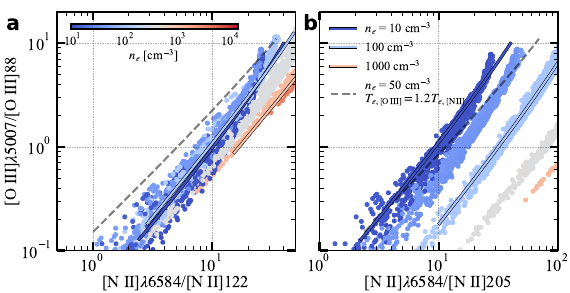}
	\caption[\teoiiifir{} vs. \teniifir{}.]{Comparison of \teoiiifir{} by \oiiio{5007}/\oiii{88} to \teniifir{} by \bsf{a}, \niio{6584}/\nii{122}; and \bsf{b} \niio{6584}/\nii{205}. The different colored lines with black edge are emissivity ratios assuming different \edens{} but temperature equilibrium, while the gray dashed thick line assumes \edens{}~=~50\,\cc{} and \teoiiifir{}~=~1.2~\teniifir{}, the same as that in \ppi{} fig.~24.}
	\label{f:model_te-te}
\end{figure}

A weak trend of \teoiiifir{} $\sim$ 1.2 \teniifir{} is reported in \ppi{}. 
In Fig.~\ref{f:model_te-te}, we plot \oiiio{5007}/\oiii{88} versus \niio{6584}/\nii{}, matching the coordinates of fig.~24 in \ppi{}. 
The model outputs, when compared to the theoretical emissivity ratios (solid colored lines with black edges), also show that \teoiiifir{} can exceed \teniifir{}, but this effect is significant only at high \te{} (\textgreater 10\textsuperscript{4}\,K). 
The models do not reproduce the observed distribution of data points around the thick gray dashed line, which would represent a systematic temperature contrast between the two ionization-state zones.

\section{Discussion}
\label{sec:model_discussion}

\subsection{Galaxies Properties and Correlations}
\label{sec:model_correlation}

In several diagnostics, particularly those related to abundances, observed line ratios exhibit much tighter correlations than predictions of aggregated model outputs. 
This suggests that, in observations, the influence of other parameters is effectively marginalized. 
There are two principal explanations for this marginalization: (1) the variation of certain parameters in real galaxies is much smaller than the range spanned in models, or (2) some physical parameters are inter-correlated. 

A clear example of minimal parameter variation is the electron density. 
Multiple diagnostics---including \nii{122/205}, \oiii{52}/\oiii{88}, \oi{145}/\cii{}, \oiii{88}/\oi{145}, \oiiio{5007}/\oio{6300}, and \teoiiifir{}---consistently indicate low electron densities, typically \edens{}~\textless~100\,\cc{}, with a representative value around 50\,\cc{}, while our models vary from 1 to 10\textsuperscript{4}\,\cc{}. 

Parameter correlations also play a significant role. 
A well-known example is the N/O--O/H relation, as a result of diverse nitrogen production channels \citep{edmunds78,pilyugin03}. 
Another important correlation involves metallicity, ionization parameter, and radiation field hardness. 
Optical line diagnostic studies, especially those utilizing the BPT diagram, have shown that low-metallicity galaxies generally have higher $U$ and harder radiation fields \citep{madden06}. 
It has been proposed that the tightness of optical line relations is a result of underlying correlations between these three properties, with metallicity being the primary driver \citep{stasinska06,vale16}. 
Accounting for the metallicity dependence of stellar winds, \citet{dopita06} derived a scaling of $q \propto Z^{-0.8}$, where $q = U \times c$ and $Z$ is the metallicity. 

To explain the dominant role of abundance in observed ratios such as \nii{}/\cii{}, a strong marginalization of other parameters is required. 
Based on the scaling relations of these lines, we infer that $U \sim (\mathrm{O/H})^{-1}$. 
Additional evidence comes from radiation field diagnostics: the joint behavior of \oiii{}/\cii{} and \neiii{}/\neii{} indicates that $U$ can increase by about an order of magnitude, and \hardness{} by approximately 0.4 dex, when moving from metal-rich U/LIRGs to dwarf galaxies with metallicities less than one-tenth solar. 
Furthermore, the $U$--abundance relation can be inferred directly from fig.~21 in \ppi{}, where \oiii{88}/\cii{} and \oiii{88}/\oi{} are compared to \nii{}/\cii{} and metallicity. 
Using the \textless 1 fit for \oiii{88}/\cii{} with the other two lines as measures of N/O, and considering the (N/O)/(O/H)~\textless~1 transition from primary to secondary production \citep[e.g.][]{henry00}, $U_1$ scales with O/H with a power-law index close to $-1$.

\subsection{Ionization Parameter and Hardness Degeneracy}
\label{sec:model_degeneracy}

Although strong intrinsic correlations exist between physical parameters, it is still important to disentangle and measure individual quantities---particularly $U$ and \hardness{}. 
This task is especially challenging because these two parameters often appear together in emission line diagnostics. 
In particular, the most widely used indicators involving \opp{} depend on the composite parameter $U_1$. 

It is important to note that $U$ and \hardness{} reflect different physical aspects: $U$ is primarily determined by the gas geometry and density structure, while \hardness{} is intrinsic to the ionizing stellar population. 
Recovering the evolution of \hardness{} with cosmic time can therefore provide key insights into the evolution of stellar populations \citep{schaerer02,kumari21}. 

To break this degeneracy, additional line ratios are required.
The \neiii{}/\neii{} ratio is particularly powerful, as it exhibits a much steeper dependence on \hardness{} and shows significant variation in observational data. 
Analysis with Fig.~\ref{f:model_OIII_CII-NeIII_NeII} demonstrates that local dwarf galaxies possess higher $U$ and higher \hardness{}, with the change in hardness being especially pronounced. 
Other line ratios that respond differently to $U$ and \hardness{} can also help disentangle these parameters. 
For example, \oiiio{5007}/\oio{6300} serves as an additional diagnostic, though it is less sensitive than \neiii{}/\neii{} and the observed dynamic range is smaller. 

In summary, while standard diagnostics are sensitive to a combination of $U$ and \hardness{}, employing multiple line ratios with different dependencies is essential to independently constrain these fundamental physical properties.

\subsection{Challenges in Photoionization Modeling}
\label{sec:model_challenge}

\subsubsection{Effect of Dust Radiation Pressure}
\label{sec:model_dust}

\begin{figure}[]
	\centering
	\includegraphics[width=\halfwdth]{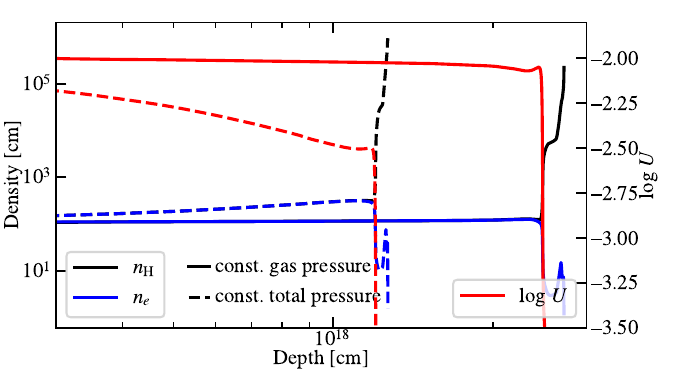}
	\caption[Model structure with dust radiation pressure. ]{Comparison of the structure of a model that uses a constant gas pressure (solid lines) with that of a model with the same setup except for adopting a constant total pressure that includes dust radiation pressure (dashed lines). In the figure, \edens{} and the total density \hdens{} are plotted on the y axis on the left and log $U$ is plotted on the right y axis, in the labeled colors. }
	\label{f:model_const_pres}
\end{figure}

As discussed in Sec.~\ref{sec:model_grid}, our photoionization models assume equilibrium based solely on gas pressure, neglecting the potential contribution from radiation pressure, particularly that exerted on dust grains. 
This modeling choice is a compromise because of the unphysical density structures produced when both radiation pressure and dust are included. 
Fig.~\ref{f:model_const_pres} demonstrates this effect by comparing two models: one assuming only constant gas pressure (solid lines), and the other incorporating radiation pressure with constant total pressure (dashed lines). 
Both models use identical parameters: plane-parallel geometry, log $U_{in}$~=~--2, initial hydrogen density $n_\mathrm{H}$~=~100\,\cc{}, \oh{}~=~--3.2, \no{}~=~--0.6, and a stellar age of 6 Myr.

The results show that, while the constant gas pressure model maintains a nearly uniform density throughout the ionized region, the inclusion of radiation pressure leads to a continuous rise in density toward the ionization front. 
As a consequence, the radiation pressure model exhibits a significantly reduced depth---less than half that of the gas pressure-only model---along with a 0.5 dex lower log $U$ and a 0.5 dex higher neutral gas column. 
These differences are not due to enhanced dust absorption of ionizing photons, as both models have identical D/G and comparable ionizing photon budgets (as indicated by similar $\int n_e^2\,\mathrm{d}V$). 
Instead, the structural differences arise from the impact of dust radiation pressure.

Near the radiation incident face of the cloud, where the photon energy flux is maximal, dust radiation pressure---proportional to the absorbed photon energy---can dominate the total pressure. 
As radiation is absorbed or converted to lower energy at larger depth, dust radiation pressure diminishes, requiring a higher gas pressure (i.e., a rise in density) to maintain pressure equilibrium. 
This effect is especially pronounced at high O/H (or high D/G) or high $U$, and can lead to density increases of up to three orders of magnitude in some models.

This non-negligible contribution of dust radiation pressure may help explain discrepancies in density diagnostics, such as the systematically lower neutral gas density inferred from \oi{145}/\cii{} compared to the values required to match observed \oi{145}/\cii{} ratios (see Fig.~\ref{f:model_n_NII-n_neut_model}). 
It also suggests a potential mismatch between the density structure in our models and that of real ISM conditions.

\begin{figure}[h]
    \centering
    \includegraphics[width=\halfwdth]{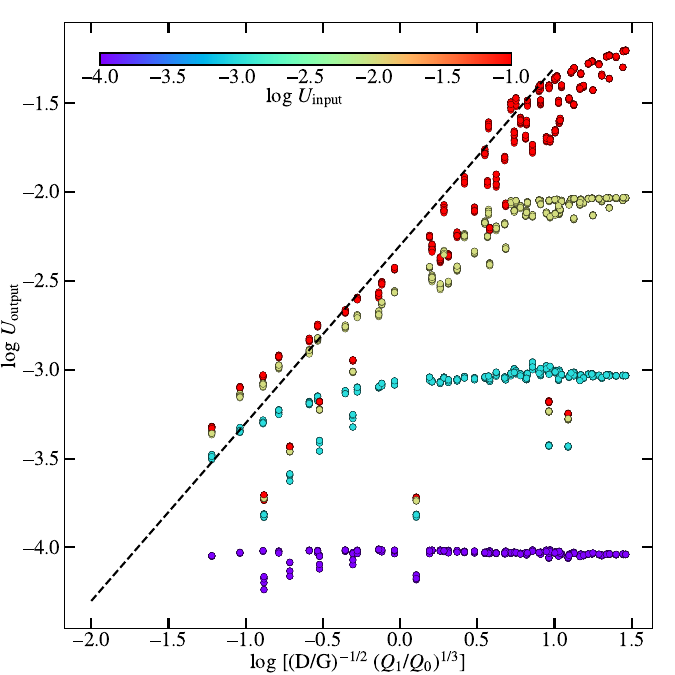}
    \caption[$U$ being limited by D/G and \hardness{} in constant pressure dusty models.]{The output ionization parameter $U$ as a function of power-law scaling of (D/G) and \hardness{}, color-coded by different model input $U$. The dashed line highlights the cap on output $U$ set by the scaling in the x axis.}
    \label{f:model_dusty_hii}
\end{figure}

This issue has previously been discussed by \citet{draine11}, who also noted that under static pressure equilibrium, dust is pushed outward and carries gas to concentrate in a thin shell near the ionization front. 
\citet{draine11} provided an analytical expression for $U$ in this regime, showing that $\log U$ is effectively limited to values below $-2$. 
In our models, which allow for a broader set of parameters, we find that the output $U$ in constant total pressure models is capped according to D/G (through O/H) and radiation hardness, following $U$ \textless 0.005 (D/G)\textsuperscript{-1/2} (\hardness{})\textsuperscript{1/3} (dashed line in Fig.~\ref{f:model_dusty_hii}). 
Thus, models with higher D/G (and hence higher dust radiation pressure), lower \hardness{}, or higher fraction of photons subject to dust absorption, are more susceptible to this limitation. 

This outcome is unphysical because, in reality, dust grains would begin to drift if radiation pressure represents a significant fraction of the total pressure. 
Additionally, the large density jumps predicted near the ionization front are not observed. 
\citet{draine11} calculated that the dust drift velocities can exceed 100 km\,s\textsuperscript{-1} in the inner regions. 
Therefore, to achieve more realistic structures, it is necessary to move beyond static photoionization models and consider grids of dusty models including both radiation pressure and dynamical evolution. 

Although Fig.~\ref{f:model_dusty_hii} and our analytical approximation presents an unphysical structure and should not be used for normalization, its scaling relations offer insight into observed correlations among \hii{} regions. 
The first is the O/H--$U$--\hardness{} correlation: newly formed massive stars create \hii{} regions with initially high $U$, but strong radiation pressure acts to redistribute dust and gas, lowering $U$ over time. 
This dynamical evolution could naturally result in higher $U$ values for low-metallicity or harder radiation field environments where dust absorption becomes inefficient. 

Another observed correlation is the size–density relation, $n_e \sim 1/D$, where $D$ is the size of the \hii{} region \citep[e.g.,][]{kennicutt84,hernandez05}. 
Follow-up studies \citep{gilbert07} find that different galaxies have distinct normalization for this relation. 
Since $U$ is defined by the ratio of ionizing photon density to electron density, and under the assumption of a radiation-bounded spherical geometry ($Q_0 \sim n_e^2 R^3$, $U \sim Q_0/(n_e R^2 c) \sim n_e R$), the observed inverse size–density relation implies a roughly constant $U$ within galaxies. 
Observationally, $n_e R$ is highest in dwarf galaxies and lowest in Galactic samples, as seen in fig.~2 of \citet{hunt09}.

\subsubsection{Problem with \oi{} Lines}

There are notable mismatches between models and observed \oi{} line ratios. 
For example, the modeled \oi{63}/\oi{145} ratio, whether intrinsic or emergent, consistently exceeds the observed value of $\sim$10 (see Fig.~\ref{f:model_n_NII-n_OI}). 
This discrepancy persists across various photoionization and PDR models \citep[e.g.,][]{kaufman06}, where \oi{63}/\oi{145} only approaches 10 under low densities (\hdens{}~\textless~10\textsuperscript{3}\,\cc{}). 
Under typical PDR conditions ($G_0/n_\mathrm{H^0} \sim 1$), the ratio remains above 13 \citep{rybak20}. 
The exact reason for this disagreement remains unclear. 

Notably, the observed \oi{63}/\oi{145} ratio shows very little scatter (0.18 dex). 
The tightness of \oi{63}/\oi{145} distribution is even more puzzling given that \oi{63} appear self-absorbed at vastly different degrees in observations of star-forming regions \citep{karska14}. 
Together, these suggest robust and universal, yet not fully understood, constraints on the conditions for global \oi{} emission.

\subsection{Limitations of Photoionization Models}

The applicability of photoionization models for interpreting observed lines must be considered with caution. 
A key assumption is that the modeled conditions adequately represent the ensemble of emitting gas, but real galaxies comprise a wide range of \hii{} region properties, with the density-size relation being one example, and the ionizing luminosity varies by up to four orders of magnitude \citep{santoro22}. 
However, the galaxy-integrated line emissions probes luminosity-averaged properties, which is dominated by the giant \hii{} regions based on the luminosity function \citep{kennicutt89,santoro22}. 

Furthermore, the origin of FIR fine-structure lines (FSL) and dust emission can be complex, particularly in the presence of AGN. 
AGN activity can significantly affect both optical and FIR line diagnostics, as confirmed by the identification of AGN contributions to \oiii{88} in \ppi{}, and the prevalence of AGN in the most IR-luminous systems. 
AGN may also cause the different FSL properties observed in high-\zz{} galaxies, which we will expand in \ppiii{}. 
Thus, comprehensive FSL diagnostics require both observations of AGN-related spectral features and photoionization model grids representative of AGN conditions.

Last but not least, we warn readers that photoionization models are not the endpoint to line diagnostics. 
While photoionization models incorporate a wide range of physical processes and offer more sophisticated predictions, they are subject to explicit and implicit assumptions and limitations in their setup and computation, and they are not necessarily more accurate in reproducing ISM properties than simple analytic arguments (like Eq.~1 in \ppi{}). 
For example, the explicit assumption that all emission arises from a single, uniform ionized structure is unrealistic for most galaxies. 
Ultimately, observations and empirical relations provide the most reliable foundation for understanding ISM properties. 
Model--observation agreement highlights the physical parameters most relevant to emission line strengths, while discrepancies are equally important, reflecting missing processes or incorrect assumptions in the models and our understandings.

\subsection{Limitations of FIR FSL Diagnostics and Studies}
\label{sec:model_limitation}

Taking a step back from detailed modeling and interpretation, it is important to recognize the fundamental limitations inherent in spectral line diagnostics. 
If $N$ independent spectral lines are observed, one can construct at most $N-1$ independent line ratios that cancel out the effect of absolute luminosity (``the bigger-things-brighter'' effect). 
The total number of pieces of information available from these independent ratios is, by definition, not greater than $N-1$.

Focusing on FIR FSLs, the eight lines discussed in this work are commonly used for studying the properties of ionized and neutral ISM. 
However, several of these lines convey redundant or degenerate information. 
For example, \nii{122/205}, \oiii{52/88}, and \ci{}/\cii{} are all primarily sensitive to the gas density. 
Similarly, \oiii{}/\cii{} and \niii{}/\nii{122} ratios are mainly driven by the ionization parameter. 
Additionally, the \oi{} doublet ratios are observed to be nearly invariant, rendering them ineffective for diagnostic purposes. 
Consequently, the number of truly independent and informative line ratios is reduced to about three. 

According to model inference, these three independent FIR FSL ratios allow determination of only three physical parameters: the helium-ionization parameter $U_1$, electron density $n_e$, and the abundance ratio N/O. 
Yet, as demonstrated in our model grid, at least five physical parameters are fundamentally relevant for characterizing the ISM: $U$, \hardness, $n_e$, N/O, and O/H. 
Each parameter is related to distinct aspects of the environment, gas distribution, or stellar population within star-forming regions. 
Thus, diagnostics based solely on FIR FSLs are inherently limited and cannot fully characterize the ISM. 
Breaking parameter degeneracies and achieving a complete physical picture requires additional information.

One approach, discussed in Sec.~\ref{sec:model_degeneracy}, is to incorporate mid-infrared (MIR) lines, which can help disentangle the $U$–\hardness{} degeneracy. 
Alternatively, including optical lines---sensitive to different parameter combinations---greatly enhances diagnostic power. 

These fundamental information limits are further compounded by observational biases. 
Current samples exhibit significant selection effects in both line detection and galaxy type, as detailed in \ppi{}. 
Critical line--galaxy combinations are underrepresented, such as \nii{} (especially 205\,\um{}) in dwarf galaxies and in LBG/LAEs, as well as intrinsically faint or wavelength-challenging lines like \niii{} and \oi{145} at all redshifts. 
Moreover, IR-bright galaxies---although representing only a small fraction of the galaxy population and showing little variation in ISM properties---dominate line detections due to their luminosity. 
On the other hand, low-mass galaxies---exhibit a wide range of metallicities that is essential for constraining key diagnostic relations---are much less luminous and have few to none detections in many FSLs. 
This bias in galaxy type potentially biases our understanding of FIR FSL distributions toward the properties of these systems.

In summary, while FIR FSLs provide powerful diagnostics, their information content is fundamentally limited by the number of independent observables, parameter degeneracies, and observational biases. 
Therefore, the future of studies on FIR FSL diagnostics does not depend solely on acquiring more FIR data, but comprehensive FIR and MIR FSL surveys dedicated to galaxy samples that are carefully assembled based on their property diversities, complemented by multiwavelength observations.

\subsection{Prospect of FIR FSL study}
\label{sec:model_prospect}

Despite their similarities to optical diagnostics and the inherent limitations in breaking certain parameter degeneracies, FIR FSLs offer several unique advantages for studying the ISM. 
The future of FIR FSL research lies in leveraging these strengths---particularly their uniqueness in diagnostic, the power of interferometric observations, and synergy with other wavelengths. 

First, FIR FSLs are unique probes for some properties in ISM. 
Line ratios like \nii{122/205} and \oi{}/\cii{} remain the only reliable tracers of low electron densities that are common to ISM in reality, as discussed in \ppi{}. 
For neutral gas, \cii{} and \oi{} are the only commonly observed tracers for the neutral atomic gas, in addition to \hi{} 21\,cm line which is still limited to very nearby universe. 
FIR FSLs are also indispensable tools for studying ISM in highly obscured systems like U/LIRGs or DSFG, where the extreme dust attenuation can make optical lines difficult or impossible to observe. 

FIR FSLs also benefit from advancements in observational technology, particularly the spatial resolution achievable with interferometers such as ALMA. 
These instruments enable sub-0.1\arcsec{} studies of the neutral and ionized gas distributions even in high-\zz{} galaxies. 
Moreover, interferometric observations naturally provide both spatial and spectral information over the entire field of view, making FSLs exceptionally powerful for investigating resolved galaxy dynamics. 
Moreover, the diagnostic power of FIR FSLs will increase significantly with a large amount of higher-fidelity data. 
If observational statistical uncertainties can be reduced below 0.1\,dex---smaller than the intrinsic scatter of model parameter fits and the typical observational systematic uncertainties---it becomes possible to study the influence of secondary parameters beyond the primary diagnostic trends.

\section{Summary}
\label{sec:model_summary}

In \ppi{} and this work, we investigate the diagnostic power of FSLs, mainly FIR FSLs, by systematically comparing galaxy integrated observational data with photoionization models. 

\begin{itemize}
    \item We run and provide a grid of photoionization models that spans five parameters and covers representative ranges of ISM conditions. 
    \item We introduce a power-law fitting approach to address the high-dimensionality of model parameter space. 
    \item We revisit and quantify the diagnostic frameworks presented in \ppi{}
\end{itemize}

Our joint analysis on observational data and models demonstrate that three key physical properties can be inferred from the FIR FSL ratios:

\begin{itemize}
	\item Ionization parameter and radiation hardness combined ($U_1$), as probed by \oiii{88}/\cii{} or \oiii{88}/\oi{145} or \niii{}/\nii{122}
	\item Electron or gas density (\edens{} or \hdens{}), as diagnosed by \nii{122/205} or \oi{145}/\cii{}
	\item Nitrogen-to-oxygen abundance ratio (N/O), by \niii{}/\oiii{88} or \nii{}/\cii{}
\end{itemize}

MIR or optical lines are needed to extract additional information, such as \hardness{}, and \te{}. 
We also highlight the importance of parameter marginalization in producing the tight empirical relations observed in \ppi{}. 
Some parameters, such as \edens{}, show limited variation in real galaxies. 
Others are marginalized due to intrinsic correlations---such as the N/O--O/H and O/H--$U$--\hardness{} relations.

We identify key bottlenecks in advancing FIR FSL diagnostics. 
On the observational side, progress is hindered by the scarcity of data for certain FIR FSLs, lack of ancillary multiwavelength observations, bias in the galaxy samples, and limited availability of MIR and optical lines at high-\zz{} that are critical for breaking parameter degeneracies and identifying AGN contributions. 

On the theoretical side, challenges remain in modeling dusty \hii{} regions that account for dynamical evolution driven by dust radiation pressure. 
The underlying physical origin and universality of the observed O/H--$U$--\hardness{} correlation, and the nature of the nearly invariant \oi{63/145} line ratio, also remain unresolved. 

Additional questions exist in understanding FIR FSLs, especially their origins in ISM and their relation to total IR luminosities. 
These questions will be explored in \ppiii{}.

At this stage, we conclude that the future of FIR FSL studies relies on (1) a deeper understanding of the physical origins of these diagnostics, which can only be achieved with a comprehensive FSL survey on diverse galaxy samples combined with multi-wavelength observations, and (2) applications that leverages their unique strengths---particularly their ability to probe low-density or highly obscured ISM, their synergy with high-resolution or high-fidelity data. 
Progress in both observational and theoretical fronts will be essential to fully realize the diagnostic potential of FIR FSLs for the study of galaxy evolution and the ISM.


\begin{acknowledgments}
We thank Iker Millan Irigoyen for the discussion on IMF. 
B.P. acknowledges the support of NRAO SOS 1519126. 
Support for this work at Cornell was provided in part by NASA grant NNX17AF37G, NSF grants AST-1716229 and AST-1910107 and NASA/SOFIA grant NNA17BF53C (SOF09-0185).  

\software{Astropy \citep{astropy13,astropy18}, 
		  \cloudy{}\footnote{Available at \url{https://trac.nublado.org/}} \citep{cloudy23,cloudy23.01}}
\end{acknowledgments}



\appendix

\section{Model Catalog}
\label{sec:model_catalog}

We provide the photoionization models used in the paper to facilitate further studies on FSLs. 
The table records integrated line intensities of many spectral lines, with a focus on connecting FIR FSLs and MIR FSLs with optical lines, and connecting different phases of gas. 

The table of the cloudy output \footnote{FLAMES-cloudy. The full machine-readable table will be available for download in the online published version. Prior to that, the catalog can be obtained by contacting the correspondence author. }is organized in the following format. 
The definitions of parameters, especially spectral line symbols, can be checked in \cloudy{} documentation \footnote{Available at \url{https://gitlab.nublado.org/cloudy/cloudy}}. 

\begin{itemize}
    \item Columns 1-5. Model setup parameters of log $U$, log \hdens{}, \no{}, \oh{}, stellar population age. 
    \item Columns 6-12. Radiation properties of the initialized models: model setup luminosity and luminosity unit; ionizing photo flux, log $\Phi_\mathrm{i}$, and log $\Phi_\mathrm{0}$ to $\Phi_\mathrm{3}$.
    \item Columns 13-24. Absolute log elemental abundance of H, He, C, N, O, Ne, Mg, Si, S, Cl, Ar, Fe. 
    \item Columns 25-27. Stellar atmosphere input setup commands. 
    \item Columns 28-31. Dust input setup commands.
    \item Columns 32-42. Model output geometries: thickness of the whole model, within \hii{} region, and within PDR (see Sec.~\ref{sec:model_structure} for definitions); \av{} at the edge of the whole model, \hii{} region front, and PDR front; total gas mass within the three zones; total H\textsuperscript{0} mass of the model; total H\textsuperscript{+} mass of the model. 
    \item Columns 43-48. Average log $U$ in different zones and weights. 
    \item Columns 49-55. Average density in different zones and weights. 
    \item Columns 56-176. Integrated line fluxes of 121 lines, including most of the strong lines from H~{\sc i}, He~{\sc i}, He~{\sc ii}, C~{\sc i}, C~{\sc ii}, C~{\sc iii}, C~{\sc iv}, N~{\sc i}, N~{\sc ii}, N~{\sc iii}, O~{\sc i}, O~{\sc ii}, O~{\sc iii}, O~{\sc iv}, Ne~{\sc ii}, Ne~{\sc iii}, Ne~{\sc iv}, Ne~{\sc v}, Ne~{\sc vi}, Na~{\sc iii}, Na~{\sc iv}, Na~{\sc vi}, Mg~{\sc ii}, Mg~{\sc iv}, Mg~{\sc v}, Mg~{\sc vii}, Mg~{\sc viii}, Si~{\sc i}, Si~{\sc ii}, Si~{\sc iii}, Si~{\sc iv}, Si~{\sc vi}, Si~{\sc vii}, Si~{\sc ix}, S~{\sc i}, S~{\sc ii}, S~{\sc iii}, S~{\sc iv}, Cl~{\sc iii}, Ar~{\sc ii}, Ar~{\sc iii}, Ar~{\sc iv}, Ar~{\sc v}, Ar~{\sc vi}, Fe~{\sc ii}, Fe~{\sc iii}, Fe~{\sc vii}. 
\end{itemize}

Table~\ref{tab:model_catalog} shows a fraction of the table as an example.

\begin{rotatetable*}
\begin{deluxetable*}{ccccccccccccccc}
\centerwidetable
    \tabletypesize{\scriptsize}
    \tablecaption{Example of the photoionization model output table. \label{tab:model_catalog}}
    \tablehead{
    \colhead{logU} & \colhead{lognH} & \colhead{log\_N\_O} & \colhead{log\_O\_H} & 
    \colhead{age} & \colhead{logPhi0} & \colhead{logPhi1} & \colhead{nH\_mean\_pdr} & \colhead{H\_\_1\_656280A} & 
    \colhead{C\_\_2\_157636M} & \colhead{N\_\_2\_121769M} & 
    \colhead{N\_\_2\_205283M} & \colhead{O\_\_3\_500684A} & 
    \colhead{O\_\_3\_883323M} }
    \startdata
-3.0 & 1.0 & -0.9 & -4.2 & 3000000.0 & 1.9431e+01 & 1.7838e+01 & 8.2332e+03 & 4.0629e-04 & 4.6012e-04 & 1.9773e-06 & 1.9796e-06 & 1.4594e-04 & 3.4570e-05 \\
-3.0 & 1.0 & -0.9 & -4.0 & 3000000.0 & 1.9446e+01 & 1.7760e+01 & 6.0875e+03 & 4.0122e-04 & 4.2798e-04 & 3.0969e-06 & 3.1428e-06 & 1.4698e-04 & 4.4344e-05 \\
-3.0 & 1.0 & -0.9 & -3.8 & 3000000.0 & 1.9475e+01 & 1.7587e+01 & 5.2783e+03 & 3.9906e-04 & 4.7574e-04 & 4.8192e-06 & 4.9538e-06 & 1.1220e-04 & 4.8871e-05 \\
-3.0 & 1.0 & -0.9 & -3.6 & 3000000.0 & 1.9504e+01 & 1.7371e+01 & 4.8132e+03 & 3.9886e-04 & 5.4880e-04 & 7.5755e-06 & 7.8533e-06 & 6.7857e-05 & 4.8656e-05 \\
-3.0 & 1.0 & -0.9 & -3.4 & 3000000.0 & 1.9528e+01 & 1.7143e+01 & 3.7130e+03 & 4.0079e-04 & 6.3975e-04 & 1.1600e-05 & 1.2010e-05 & 3.4004e-05 & 4.8266e-05 \\
-3.0 & 1.0 & -0.9 & -3.2 & 3000000.0 & 1.9515e+01 & 1.7274e+01 & 2.6965e+03 & 3.9852e-04 & 6.9896e-04 & 1.6016e-05 & 1.6237e-05 & 2.1256e-05 & 8.3939e-05 \\
-3.0 & 1.0 & -0.9 & -3.0 & 3000000.0 & 1.9455e+01 & 1.7709e+01 & 1.4180e+03 & 3.9434e-04 & 6.9956e-04 & 2.0644e-05 & 2.1075e-05 & 7.1316e-06 & 1.9542e-04 \\
-3.0 & 2.0 & -0.9 & -4.2 & 3000000.0 & 2.1733e+01 & 2.0141e+01 & 5.4630e+04 & 4.1239e-03 & 1.7160e-03 & 1.5711e-05 & 4.7045e-06 & 1.4796e-03 & 2.6107e-04 \\
-3.0 & 2.0 & -0.9 & -4.0 & 3000000.0 & 2.1748e+01 & 2.0062e+01 & 4.4435e+04 & 4.0505e-03 & 1.7368e-03 & 2.4774e-05 & 7.4316e-06 & 1.4965e-03 & 3.3134e-04 \\
-3.0 & 2.0 & -0.9 & -3.8 & 3000000.0 & 2.1777e+01 & 1.9890e+01 & 3.7393e+04 & 4.0440e-03 & 1.8764e-03 & 3.9286e-05 & 1.1674e-05 & 1.1712e-03 & 3.6387e-04 \\
-3.0 & 2.0 & -0.9 & -3.6 & 3000000.0 & 2.1806e+01 & 1.9673e+01 & 3.0922e+04 & 4.0341e-03 & 2.0741e-03 & 6.2206e-05 & 1.8045e-05 & 7.3512e-04 & 3.5915e-04 \\
-3.0 & 2.0 & -0.9 & -3.4 & 3000000.0 & 2.1830e+01 & 1.9445e+01 & 2.3025e+04 & 4.0350e-03 & 2.3020e-03 & 9.5163e-05 & 2.6696e-05 & 4.0457e-04 & 3.5545e-04 \\
-3.0 & 2.0 & -0.9 & -3.2 & 3000000.0 & 2.1817e+01 & 1.9577e+01 & 1.6525e+04 & 3.9659e-03 & 2.4582e-03 & 1.2886e-04 & 3.4720e-05 & 3.2755e-04 & 6.2552e-04 \\
-3.0 & 2.0 & -0.9 & -3.0 & 3000000.0 & 2.1758e+01 & 2.0012e+01 & 1.0167e+04 & 3.8611e-03 & 2.5143e-03 & 1.6649e-04 & 4.4336e-05 & 2.1735e-04 & 1.4603e-03 \\
-2.0 & 1.0 & -0.9 & -4.2 & 3000000.0 & 2.1733e+01 & 2.0141e+01 & 5.9482e+03 & 3.3896e-03 & 1.1948e-03 & 2.7580e-06 & 2.7379e-06 & 4.6729e-03 & 8.7917e-04 \\
-2.0 & 1.0 & -0.9 & -4.0 & 3000000.0 & 2.1748e+01 & 2.0062e+01 & 4.2051e+03 & 3.1465e-03 & 1.1439e-03 & 4.2180e-06 & 4.2883e-06 & 4.7898e-03 & 1.1814e-03 \\
-2.0 & 1.0 & -0.9 & -3.8 & 3000000.0 & 2.1777e+01 & 1.9890e+01 & 3.1868e+03 & 2.8984e-03 & 1.1647e-03 & 6.8131e-06 & 7.0579e-06 & 4.0629e-03 & 1.4707e-03 \\
-2.0 & 1.0 & -0.9 & -3.6 & 3000000.0 & 2.1806e+01 & 1.9673e+01 & 2.7841e+03 & 2.6386e-03 & 1.2071e-03 & 1.2258e-05 & 1.2830e-05 & 2.8806e-03 & 1.6999e-03 \\
-2.0 & 1.0 & -0.9 & -3.4 & 3000000.0 & 2.1830e+01 & 1.9445e+01 & 2.2981e+03 & 2.3766e-03 & 1.2734e-03 & 2.2735e-05 & 2.3915e-05 & 1.7396e-03 & 1.8713e-03 \\
-2.0 & 1.0 & -0.9 & -3.2 & 3000000.0 & 2.1817e+01 & 1.9577e+01 & 1.5544e+03 & 2.1315e-03 & 1.2469e-03 & 2.8486e-05 & 2.9954e-05 & 1.2304e-03 & 2.6157e-03 \\
-2.0 & 1.0 & -0.9 & -3.0 & 3000000.0 & 2.1758e+01 & 2.0012e+01 & 8.9931e+02 & 1.9674e-03 & 1.1607e-03 & 3.2535e-05 & 3.3926e-05 & 7.7748e-04 & 4.0288e-03 \\
-2.0 & 2.0 & -0.9 & -4.2 & 3000000.0 & 2.4036e+01 & 2.2443e+01 & 4.2141e+04 & 3.4102e-02 & 2.9215e-03 & 2.2257e-05 & 6.8370e-06 & 4.7298e-02 & 6.4995e-03 \\
-2.0 & 2.0 & -0.9 & -4.0 & 3000000.0 & 2.4051e+01 & 2.2365e+01 & 3.2223e+04 & 3.1594e-02 & 2.9765e-03 & 3.4814e-05 & 1.0885e-05 & 4.8945e-02 & 8.5773e-03 \\
-2.0 & 2.0 & -0.9 & -3.8 & 3000000.0 & 2.4080e+01 & 2.2192e+01 & 2.4754e+04 & 2.9024e-02 & 3.1304e-03 & 5.6917e-05 & 1.7788e-05 & 4.2254e-02 & 1.0447e-02 \\
-2.0 & 2.0 & -0.9 & -3.6 & 3000000.0 & 2.4109e+01 & 2.1976e+01 & 2.0223e+04 & 2.6316e-02 & 3.3213e-03 & 1.0217e-04 & 3.1082e-05 & 3.0947e-02 & 1.1816e-02 \\
-2.0 & 2.0 & -0.9 & -3.4 & 3000000.0 & 2.4133e+01 & 2.1748e+01 & 1.6208e+04 & 2.3514e-02 & 3.5494e-03 & 1.8750e-04 & 5.5229e-05 & 1.9868e-02 & 1.2766e-02 \\
-2.0 & 2.0 & -0.9 & -3.2 & 3000000.0 & 2.4120e+01 & 2.1879e+01 & 1.1429e+04 & 2.0763e-02 & 3.5624e-03 & 2.3198e-04 & 6.6847e-05 & 1.5877e-02 & 1.7165e-02 \\
-2.0 & 2.0 & -0.9 & -3.0 & 3000000.0 & 2.4060e+01 & 2.2314e+01 & 7.4504e+03 & 1.8359e-02 & 3.3727e-03 & 2.5490e-04 & 7.2981e-05 & 1.3389e-02 & 2.4792e-02 \\

    \enddata
\end{deluxetable*}
\end{rotatetable*}


\bibliographystyle{aasjournal}
\bibliography{main,flames-low,flames-high}{}


\end{document}